\documentclass[sigconf]{acmart}
\AtBeginDocument{%
  }

\setcopyright{none}
\renewcommand\footnotetextcopyrightpermission[1]{}
\acmConference[WSDM '27]{The 20th ACM International Conference on Web Search and Data Mining}{February 15--19, 2027}{Hong Kong, China}
\acmYear{2027}

\usepackage{enumitem}
\usepackage{pifont}
\usepackage{algorithm}
\usepackage{algpseudocode}
\usepackage{array}
\usepackage{multirow}
\usepackage{xcolor}
\usepackage{tikz}
\usepackage[most]{tcolorbox}

\DeclareRobustCommand{\glowtext}[2]{%
  \begin{tikzpicture}[baseline=(X.base)]
    \foreach \glowradius/\glowopacity in {
      1.8/0.04,
      1.4/0.06,
      1.0/0.08,
      0.6/0.12
    }{%
      \foreach \glowangle in {0,30,...,330}{%
        \node[
          inner sep=0pt,
          text=#1,
          opacity=\glowopacity
        ] at (\glowangle:\glowradius pt) {#2};%
      }%
    }%

    \node[
      inner sep=0pt,
      text=black
    ] (X) at (0,0) {#2};
  \end{tikzpicture}%
}
\definecolor{myGreen}{HTML}{CBE3C9}
\definecolor{myBlue}{HTML}{729AB0}
\definecolor{winnerred}{RGB}{255, 204, 204}
\definecolor{secondwinneryellow}{RGB}{255, 230, 153}
\newcommand{\winner}[1]{%
  \begingroup
  \setlength{\fboxsep}{0.35pt}%
  \colorbox{winnerred}{#1}%
  \endgroup
}
\newcommand{\secondwinner}[1]{%
  \begingroup
  \setlength{\fboxsep}{0.35pt}%
  \colorbox{secondwinneryellow}{#1}%
  \endgroup
}
\DeclareRobustCommand{\winnerlegend}{%
  \begingroup
  \setlength{\fboxsep}{0.35pt}%
  \colorbox{winnerred}{light-red background}%
  \endgroup
}
\DeclareRobustCommand{\secondwinnerlegend}{%
  \begingroup
  \setlength{\fboxsep}{0.35pt}%
  \colorbox{secondwinneryellow}{light-yellow background}%
  \endgroup
}
\newif\ifcompactauthors
\compactauthorstrue
\ifcompactauthors
\makeatletter
\def\@mkauthors@iii{%
  \gdef\@currentauthors{}%
  \gdef\@currentaffiliation{}%
  \global\setbox\mktitle@bx=\vbox{%
    \noindent\unvbox\mktitle@bx\par\medskip
    \centering
    {\large\normalfont
      Wenrui Zhou\textsuperscript{1}\quad
      Jiashun Liu\textsuperscript{1}\quad
      Sujie Hu\textsuperscript{2}\quad
      Bang Liu\textsuperscript{3}\quad
      Ling Pan\textsuperscript{1,*}\par}
    \smallskip
    {\normalsize\normalfont
      \textsuperscript{1}HKUST\quad
      \textsuperscript{2}Tsinghua University\quad
      \textsuperscript{3}Mila-Quebec Artificial Intelligence Institute \& Universit\'e de Montr\'eal\par}
    \bigskip}}
\makeatother
\fi

\begin{document}

\title{Beyond Code and Pixels: Co-Generating Editable and Visually Rich Presentations}

\author{Wenrui Zhou}
\affiliation{%
  \institution{The Hong Kong University of Science and Technology}
  \city{Hong Kong}
  \country{China}}
\email{wzhouav@connect.ust.hk}

\author{Jiashun Liu}
\affiliation{%
  \institution{The Hong Kong University of Science and Technology}
  \city{Hong Kong}
  \country{China}}
\email{jliuhc@connect.ust.hk}

\author{Sujie Hu}
\affiliation{%
  \institution{Tsinghua Shenzhen International Graduate School, Tsinghua University}
  \city{Shenzhen}
  \country{China}}
\email{husj24@mails.tsinghua.edu.cn}

\author{Bang Liu}
\affiliation{%
  \institution{Mila}
  \city{Montr\'eal}
  \country{Canada}}
\affiliation{%
  \institution{Universit\'e de Montr\'eal}
  \city{Montr\'eal}
  \country{Canada}}
\email{bang.liu@umontreal.ca}

\author{Ling Pan}
\authornote{Corresponding author: Ling Pan. Emails: \href{mailto:wzhouav@connect.ust.hk}{\nolinkurl{wzhouav@connect.ust.hk}} (Wenrui Zhou); \href{mailto:lingpan@ust.hk}{\nolinkurl{lingpan@ust.hk}} (Ling Pan).}
\affiliation{%
  \institution{The Hong Kong University of Science and Technology}
  \city{Hong Kong}
  \country{China}}
\email{lingpan@ust.hk}

\renewcommand{\shortauthors}{Zhou et al.}

\begin{abstract}
Generating PPTX presentations is a structured task that requires semantic planning, visual composition, precise layout, and editability, motivating collaboration among specialized agents. 
However, programmatic methods offer editability but often produce rigid designs, while image-based methods provide visual richness but flatten content into difficult-to-edit pixels. 
A hybrid system can combine these strengths, but must coordinate interdependent decisions across content, layout, and appearance. 
To address this challenge, we propose \textsc{SlidesWeaver}, a hybrid multi-agent system that coordinates language and vision agents through shared deck- and slide-level states. 
Specifically, it uses staged multimodal commitment, diffusion-based composition planning, and two generation schedules for different priorities between composition freedom and accurate text preservation, while preserving native text and geometry.
Across 1,728 generations spanning 24 system-model configurations, \textsc{SlidesWeaver} achieves consistently superior overall quality across all evaluated dimensions while maintaining structural control, editability, and practical generation efficiency.
Code and appendix: \href{https://github.com/William030422/WSDM27-SlidesWeaver}{\nolinkurl{github.com/William030422/WSDM27-SlidesWeaver}}.
\end{abstract}

\begin{teaserfigure}
  \centering
 \includegraphics[width=0.9\textwidth]{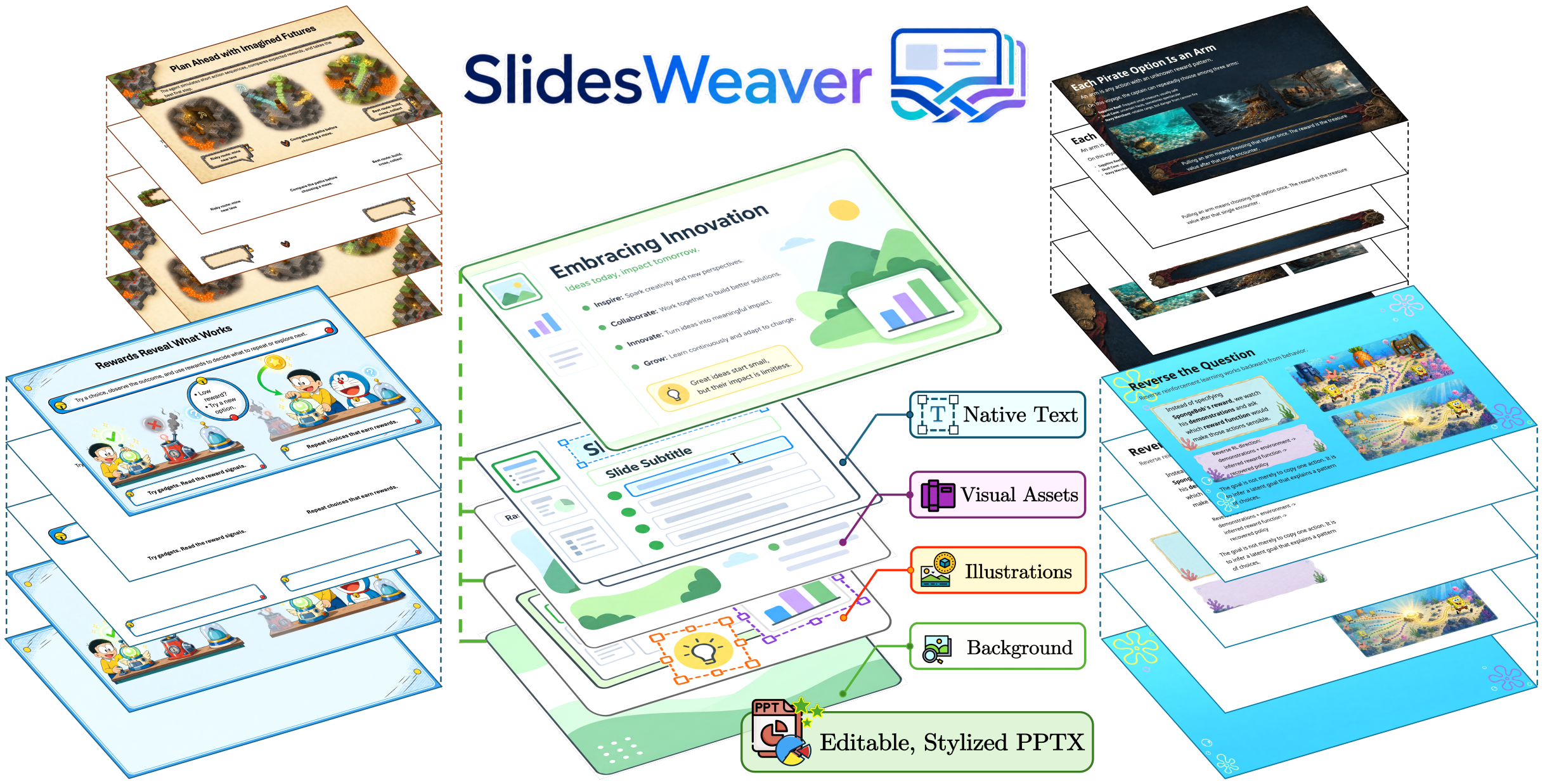}
  \caption{
    Representative slides generated by \textsc{SlidesWeaver} in four visual styles: 
    \textbf{Left} two examples are made with Topology First route with Doraemon and Minecraft styles; 
    \textbf{Right} two examples are made with Text First route with Pirates of the Caribbean and SpongeBob SquarePants styles.}
  \label{fig:teaser}
\end{teaserfigure}

\maketitle
\hypersetup{
  pdftitle={Beyond Code and Pixels: Co-Generating Editable and Visually Rich Presentations},
  pdfauthor={Wenrui Zhou, Jiashun Liu, Sujie Hu, Bang Liu, Ling Pan}
}

\section{Introduction}

Presentation generation transforms a structured presentation request together with a style brief into a coherent sequence of slides that communicates a narrative through text and visual composition \cite{fu2022doc2ppt, zheng2025pptagent, ozden2026arcdeck}.
Accordingly, a generated presentation must coordinate semantic content, visual organization, and editable structure, because these components jointly determine what the audience understands and how the deck can be revised.
Given different presentation scenarios place different weights on visual expressiveness and editability, meeting such flexible goals therefore requires an intelligent system that coordinates structural control and visual expressiveness throughout the generation process \cite{zeng2026slidetailor, chen2026xslides}.

Existing approaches largely divide into two paradigms with complementary strengths.
Programmatic methods construct slides through structured operations or executable representations, such as editing actions, layout programs, or HTML-like code \cite{ge2025autopresent, cui2026deepslides}.
Recent agentic systems organize this structured generation process across specialized roles for research, planning, and design \cite{zheng2025pptagent, zheng2026deeppresenter, liang2026slidegen, vats2026slidelab}.
While available, image generation often serves as an optional action for acquiring visual material alongside web retrieval, leaving the core composition process to the structured authoring agents.
These explicit representations provide precise control over text and geometry, yet their structural primitives tend to produce rigid or repetitive compositions with limited visual expressiveness \cite{cui2026deepslides, yang2026slidesgen}.
Conversely, image-centric methods draw on generative visual priors to create richer illustrations, styles, and compositions \cite{yang2026slidesgen}.
However, rasterization flattens text, geometry, and their relationships into pixels, making individual elements difficult to edit or recover reliably \cite{cao2026drawai,hu2026blame,zheng2026slideforge}.
Thus, current systems typically gain either structural control or visual freedom at the expense of the other.

Combining these paradigms within a multi-agent system introduces a coordination problem, which is that specialized agents make interdependent decisions through different representations.
Content, composition, and geometry constrain one another: text length determines the space required by a layout, visual composition determines how that text is grouped and perceived, and the available geometry determines how much content can be expressed clearly.
Premature commitment by any agent propagates constraints downstream. 
Fixed geometry limits subsequent visual exploration, while in contrast, a complete raster rendering removes the structure required for precise editing \cite{cui2026deepslides, cao2026drawai}.
The central question is therefore \textbf{which agent should commit content, composition, geometry, and visual appearance, and when?}
We formulate presentation generation as \emph{staged multimodal commitment}, a coordination principle through which agents progressively commit these decisions as their constraints become available.

Based on this formulation, we introduce \textsc{SlidesWeaver}, a hybrid multi-agent system that coordinates language and vision agents through shared deck- and slide-level states.
A deck-level director establishes the narrative and visual language; for each slide, specialized agents use a low-step diffusion model to explore compositions, ground a selected proposal into explicit text and visual regions, synthesize geometry-registered visual elements, and critique the rendered result.
This design places generative visual models inside both the planning and synthesis loops while retaining the structured state required for editable assembly.
Within this mechanism, \textsc{SlidesWeaver} supports two commitment schedules that prioritize different relationships between composition and text.
Figure~\ref{fig:teaser} shows representative outputs from both schedules across four distinct visual styles.
\ding{182} \emph{\textbf{Layout-conditioned text generation}} grounds a diffusion proposal before generating copy according to the resulting regions and capacities, allowing composition to shape the final wording.
\ding{183} \emph{\textbf{Text-constrained layout generation}} commits the generated text first and then grounds the diffusion proposal around that exact copy, favoring scenarios where preserving accurate, pre-committed text is a stronger requirement.
Both schedules yield the same structured PPTX representation with native text, explicit geometry, and separate control of raster elements.

Because both schedules produce the same structured PPTX representation, we evaluate \textsc{SlidesWeaver} on 72 cases spanning six domains and two complementary input tracks.
We compare it with three academic systems using visual expressiveness, instruction following, content quality, layout validity, topological diversity, cost, latency, and first-pass success as evaluation criteria.
Across 15 language-model--metric comparisons, one of our two routes achieves every best quality score, while the Text First route improves over the strongest baseline by 14.2\% in visual expressiveness, 15.9\% in instruction following, 17.7\% in content quality, 20.3\% in layout validity, and 22.9\% in topological diversity.
The slide-separable architecture also supports parallel execution and bounded visual-synthesis budgets: Topology First costs 37.3\% less and runs 57.7\% faster than the image-enabled DeepPresenter baseline, and six of the eight non-dominated configurations on the quality-cost-latency frontier belong to \textsc{SlidesWeaver}.
Our ablations support the mechanism behind these gains: removing diffusion lowers the average Topo-Vendi score from 0.84 to 0.54, while removing the layout judge lowers average Layout from 0.95 to 0.85.
Pixel-space exploration, structured grounding, and staged multi-agent coordination thus offer complementary benefits for editable presentation generation.

We summarize our contributions as follows:
\begin{itemize}[leftmargin=*]
    \item We propose \textsc{SlidesWeaver}, a hybrid multi-agent framework for editable presentation generation with two complementary routes: Topology First for layout-conditioned text generation and Text First for text-constrained layout generation.
    \item We integrate pixel-space diffusion into the PPTX pipeline to propose slide topologies and synthesize geometry-registered visual elements, enabling visually rich and topologically diverse slides with native text and editable geometry.
    \item We design deck- and slide-level multi-agent coordination that uses shared context, parallel execution, style reuse, and consolidated synthesis and achieves superior performance and strong trade-offs among quality, cost, and latency.
\end{itemize}

\section{Related Work}

\begin{figure}[!ht]
    \vspace{-5pt}
    \centering
    \includegraphics[width=\columnwidth]{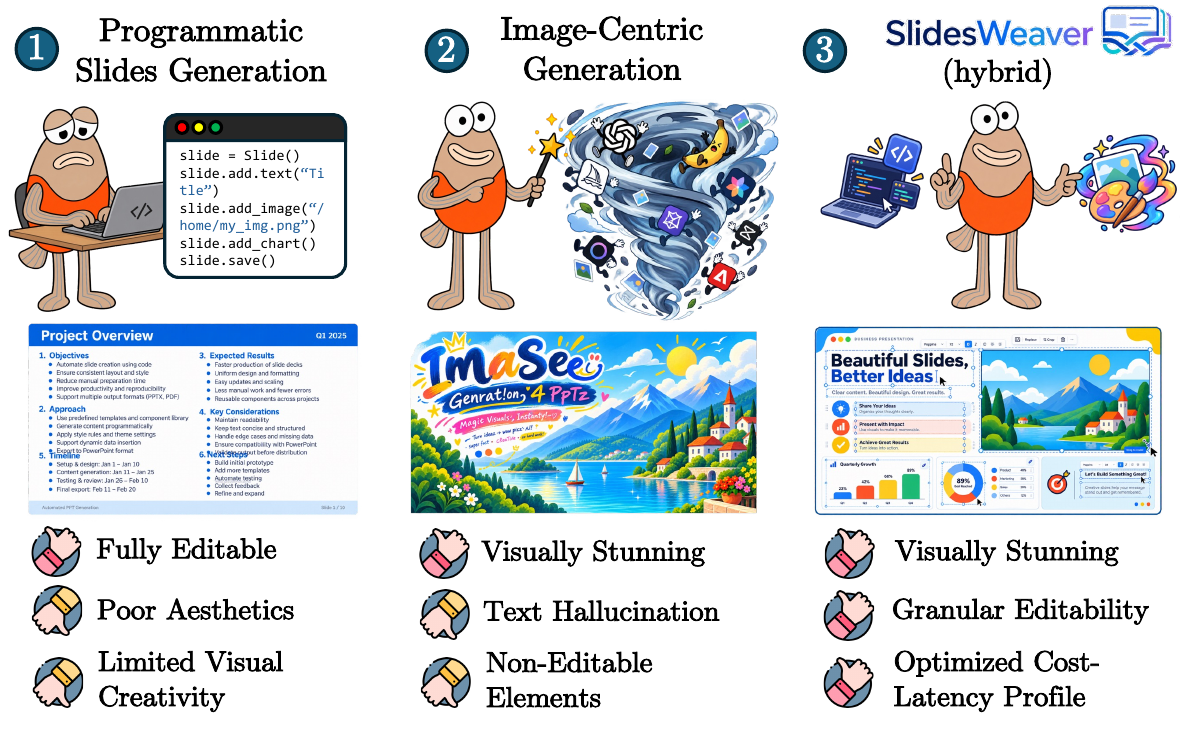}
    \vspace{-8pt}
    \caption{
    Comparison of three presentation-generation paradigms. 
    Programmatic methods preserve editability but offer limited visual creativity, while image-centric generation improves aesthetics at the cost of text fidelity and editability. 
    \textsc{SlidesWeaver} combines both strengths, producing visually rich, granularly editable slides with an optimized cost-latency profile.
    }
    \label{fig:related-work}
    \vspace{-12pt}
\end{figure}

\begin{figure*}[!t]
    \centering
    \includegraphics[width=0.9\textwidth]{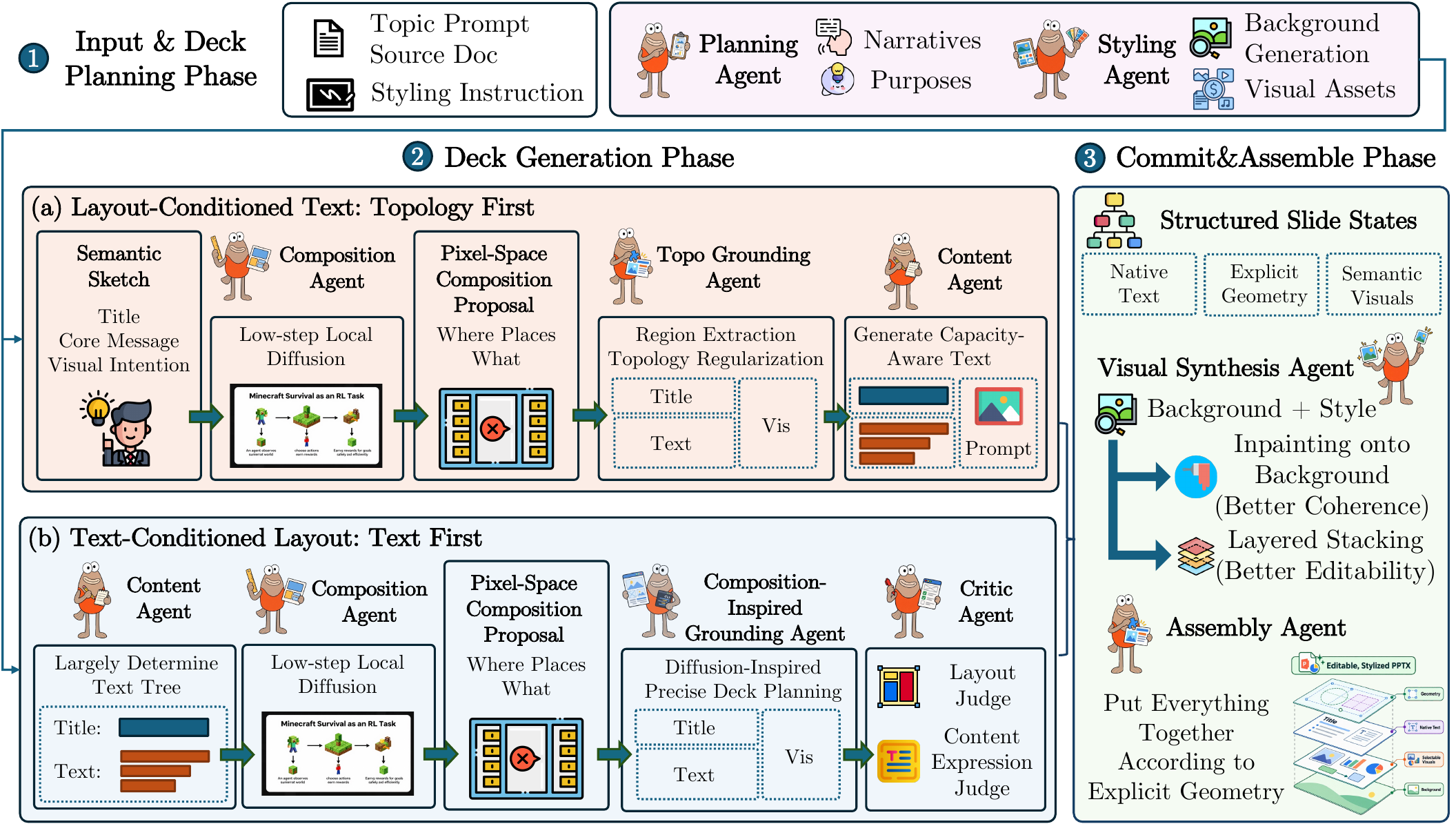}
    \vspace{-4pt}
    \caption{
    Overview of \textsc{SlidesWeaver}, a multi-agent system for hybrid PPTX generation.
    A shared Input \& Deck Planning Phase establishes deck-level semantic, narrative, and visual context.
    During Deck Generation, specialized agents follow two complementary routes: \emph{Topology First} grounds diffusion-proposed structure and adapts text to it, while \emph{Text First} preserves intended content and uses diffusion to guide and validate executable geometry.
    Finally, the Visual Synthesis and Assembly Agents convert structured slide states into coherent visual layers and editable PPTX slides.
    }
    \label{fig:method-overview}
\end{figure*}

\noindent \paragraph{\textbf{Programmatic Presentation Generation.}}
Early document-to-slide systems formulate presentation generation as structured prediction: \textsc{Doc2PPT} summarizes source documents, retrieves text and figures, and predicts their slide layouts \cite{fu2022doc2ppt}.
Recent systems replace task-specific predictors with executable representations \cite{xu2025pregenie, yang2025autoslides, fang2026seaslides}.
\textsc{AutoPresent} generates visual programs through a high-level slide API \cite{ge2025autopresent}, while \textsc{PPTAgent} preserves reference designs by producing executable editing actions \cite{zheng2025pptagent}.
Subsequent work expands the design and revision process~\cite{tian2026redeck}: \textsc{DeepSlides} separates page design from code implementation \cite{cui2026deepslides}, and \textsc{DeepPresenter} organizes generation into a researcher--presenter pipeline in which one agent prepares a manuscript and assets, another authors HTML slides, and rendered artifacts provide feedback for revision \cite{zheng2026deeppresenter}.
This progression improves planning flexibility and robustness while retaining explicit, editable structure, but the final composition remains encoded through executable primitives.

\noindent \paragraph{\textbf{Image-Centric and Editable Visual Generation.}}
Image-centric generation moves composition into pixel space and avoids restricting every visual element to a predefined slide primitive.
Current state-of-the-art image models include GPT Image 2 and Nano Banana Pro \cite{chang2026fepbench}.
The resulting renderings, however, are static raster artifacts without object separation or text reflow \cite{yang2026slidesgen, chen2026slide2ppt}.
\textsc{DrawAI} follows a top-down reconstruction route that recovers editable artifacts from completed raster images: it first infers authoring structure from flattened pixels and then synthesizes valid, manipulable objects \cite{cao2026drawai}.
This post-hoc recovery introduces an additional parsing-and-reconstruction burden because region and text proposals can be incomplete, duplicated, or semantically fragmented \cite{cao2026drawai}.
These complementary limitations call for a hybrid pipeline that uses image-generation priors for visual exploration while committing text, geometry, and object structure in editable form.


\section{Method}

\textsc{SlidesWeaver} is a hybrid visual-rich presentation generation system that combines pixel-space generative priors with structured slide representations.
Its design is motivated by a fundamental representation mismatch: visual composition is ultimately judged in pixel space, whereas an editable presentation must be realized through explicit topology, geometry, and object structure. 
We therefore separate composition proposal from structural commitment, using diffusion to propose visual organization and grounding to determine how strongly that proposal constrains the executable slide.
Figure~\ref{fig:method-overview} provides an overview of the resulting three-stage pipeline and the two complementary generation routes.

\subsection{Task Formulation}

For a structured presentation request $q$ and a style instruction $s$, our task is to generate an editable slide deck
$\mathcal{Y}=\{S_i\}_{i=1}^{n}$.
The request $q$ specifies the topic, target audience, communication objective, success criterion, language, and requested number of slides.
Each slide is associated with a semantic intent $M_i$ specifying the message and required information to preserve, and is realized as
\begin{equation}
    S_i=(X_i,T_i,G_i,A_i),
\end{equation}
where $X_i$ denotes the realized textual content, $T_i$ the relational organization among semantic regions, $G_i$ their explicit geometry, and $A_i$ the generated visual assets.
We distinguish topology from geometry because the former specifies how regions are organized, whereas the latter specifies their exact positions and sizes.
The central challenge is to realize these components jointly without sacrificing either visual freedom or structural control.

\subsection{Input \& Deck Planning Phase}

The Input \& Deck Planning Phase establishes the deck-level narrative and visual identity before generating individual slides.
The Planning Agent organizes the request $q$ into a narrative $N$ and a semantic intent $M_i$ for each slide, while the Styling Agent uses the style instruction $s$ to define a shared visual language $V$, a canonical background $B$, and reusable visual resources $R$:
\begin{equation}
    \mathcal{D}=\left(N,\{M_i\}_{i=1}^{n},V,B,R\right).
\end{equation}
These shared decisions remain deliberately high-level.
They constrain the deck's narrative and visual identity without prescribing slide-specific composition or geometry. 
Committing finer-grained layout decisions at this stage would prematurely constrain the downstream search space and reduce the room for slide-specific visual exploration.
Each slide therefore retains the flexibility to determine its own composition, text realization, and geometry in the subsequent generation phase.

\subsection{Deck Generation Phase}

\noindent \paragraph{\textbf{Why a pixel-space proposal is needed.}}
Pure code-based geometry planning (e.g. SVG-based and HTML-based) conflates composition discovery with layout execution.
In our HTML-only observations, structurally valid plans repeatedly concentrate on a small set of regular composition families.
The representation-feedback mismatch explains this behavior: balance, hierarchy, negative space, and topology become visible after rendering, while the planner must already commit the decomposition and coordinates that produce them in a vast search space.
While rendering closes the loop for correction only after the composition has been chosen, the Judge Agent can consequently repair local overflow, overlap, and spacing defects, but replacing the composition with a fundamentally different topology requires a new proposal.

\begin{figure*}[t]
\centering
\includegraphics[width=0.9\textwidth]{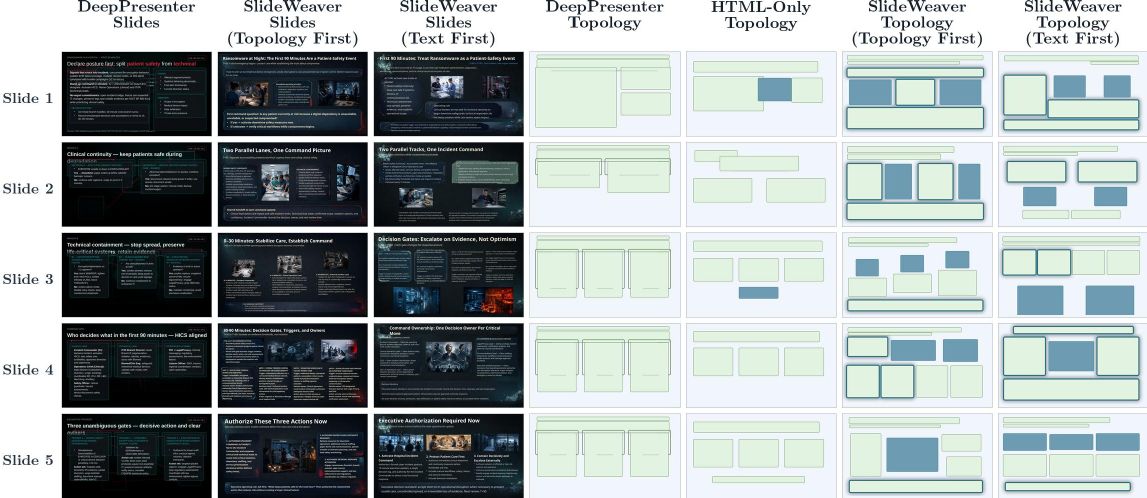}
\caption{
Qualitative comparison of topological diversity on the same five-slide hospital-ransomware brief. We apply GPT-5.4 uniformly for all methods regarding text generation and deck planning.
    The \textcolor{myGreen}{green regions} denote text, the \textcolor{myBlue}{blue regions} denote visual boxes, and a \texorpdfstring{\glowtext{myBlue}{glowing outline}}{glowing outline} marks text associated with a raster holder. 
The HTML-only ablation removes the diffusion proposal only from the Composition-Inspired Grounding Agent, resulting in similarly regular compositions to DeepPresenter, while both \textsc{SlidesWeaver} routes have more varied relational structures and higher normalized topology Vendi scores~\cite{friedman2022vendiscore}.
}
\label{fig:pixel-space-diversity}
\end{figure*}

Figure~\ref{fig:pixel-space-diversity} illustrates the effect of pixel-space proposals on a fixed five-slide hospital-ransomware brief. 
Without a diffusion proposal, the Composition-Inspired Grounding Agent produces valid but visually conservative topologies, closely resembling the regular layouts generated by DeepPresenter. 
In contrast, both \textsc{SlidesWeaver} routes explore substantially different relational organizations across the same deck, including dual-track, branching, ownership-centered, and action-triad compositions.
This qualitative difference is reflected by the normalized topology Vendi score~\cite{friedman2022vendiscore}.
While DeepPresenter obtains $0.5423$ and the HTML-only ablation drops to $0.5066$, Topology First and Text First reach $0.7986$ and $0.8042$, respectively, showing the effectiveness of pixel-space proposals in discovering diverse relational organizations.

Notably, diffusion serves two complementary roles in our framework. 
Before structural commitment, it acts as a pixel-space exploration mechanism, using its generative prior to discover creative compositions beyond predefined regions and layout primitives, including unconventional grouping, asymmetry, negative space, and visual hierarchy.
After the slide structure has been determined, diffusion instead acts as a visual synthesis mechanism, transforming the committed semantic and geometric structure into visually rich and instructive content. 
We therefore use lightweight local diffusion for open-ended composition exploration and reserve more capable remote generation for final visual realization.
Further discussion related to local diffusion is placed in Appendix~\ref{app:local-diffusion}.

\noindent \paragraph{\textbf{Topology First: layout-conditioned text.}}
Topology First method applies when the slide semantics are fixed but their exact verbal realization can adapt to the discovered composition for more coherent content delivery. 
Given the slide-level semantic intent $M_i$, the Composition Agent first generates a pixel-space proposal $P_i$ that explores a candidate visual organization. 
Rather than treating its pixel coordinates as exact geometry, the Topology Grounding Agent extracts the underlying relational structure and converts it into an executable topology together with a coarse spatial scaffold and regional capacity constraints:
\begin{equation}
(T_i,\bar{G}_i,K_i)=\Gamma_{\mathrm{topo}}(P_i).
\end{equation}
Here, $T_i$ captures the committed relational organization, while $\bar{G}_i$ provides only a soft geometric scaffold and $K_i$ describes the available capacity of each semantic region.

The Content Agent then realizes the slide text conditioned on the grounded topology and its capacities:
\begin{equation}
X_i=\Gamma_{\mathrm{content}}(M_i,T_i,K_i).
\end{equation}
The resulting dependency is
\begin{equation}
M_i \rightarrow P_i \rightarrow (T_i,\bar{G}_i,K_i) \rightarrow X_i.
\end{equation}
By postponing exact wording until the composition has been grounded, the system allows visual organization to lead content realization while preserving the original semantic intent.
This capacity-aware realization is what makes a diffusion-proposed topology executable despite the lack of precise geometric guarantees in pixel space.

\noindent \paragraph{\textbf{Text First: text-conditioned layout.}}
Text First applies when preserving accurate, pre-committed text is a stronger requirement and the wording therefore cannot freely adapt to the discovered composition.
The Content Agent first commits the verbal realization of the slide,
\begin{equation}
X_i=\Gamma_{\mathrm{text}}(q,M_i).
\end{equation}
The Composition Agent then reduces $X_i$ to a deliberately lightweight semantic description, retaining only the title, high-level communicative intent, and concise expression logic.
Detailed body content is withheld from the diffusion process so that the proposal remains weakly constrained and can freely explore alternative grouping, hierarchy, asymmetry, and visual rhythm. 
The resulting pixel-space proposal $P_i$ therefore acts as creative compositional inspiration rather than a specification of the final layout.

The Composition-Inspired Grounding Agent then reconciles the diffusion proposal with the committed text and produces an executable HTML plan,
\begin{equation}
H_i^{(0)}=\Gamma_{\mathrm{guide}}(X_i,P_i).
\end{equation}
Unlike DeepPresenter, which generates HTML slides and subsequently converts them into PowerPoint, we use HTML only as an intermediate planning representation~\cite{zheng2026deeppresenter}.
We take advantages of HTML-based topology planning given its explicit structure and browser-verifiable rendering to provide a convenient geometric contract for grounding the otherwise free-form visual proposal, while the final slide is assembled separately from editable text and generated visual assets.

Importantly, the diffusion proposal remains advisory rather than binding in this route. 
The grounding agent may revise its suggested organization whenever necessary to preserve the committed content and obtain a valid structured layout. 
Browser rendering provides perceptual evidence for refinement, whereas the HTML structure provides authoritative region geometry:
\begin{equation}
I_i=\rho(H_i),
\qquad
(T_i,G_i)=\gamma(H_i).
\end{equation}

Given its verifiable nature, a Judge Agent iteratively evaluates the rendered layout and repairs structural defects such as overflow, overlap, or inconsistent text-visual organization without changing the committed content. 
The final, refined HTML plan $H_i^{\star}$ defines the topology and metric geometry for downstream synthesis:
\begin{equation}
X_i \rightarrow P_i \rightarrow H_i^{(0)}
\xrightarrow{\mathrm{refinement}} H_i^{\star}
\xrightarrow{\gamma} (T_i,G_i).
\end{equation}
Thus, Text First uses diffusion to broaden the compositions considered by the structured planner, while retaining HTML as the final authority over content feasibility and exact geometry.

\subsection{Commit \& Assemble Phase}

After grounding, the two routes differ in how strongly geometry has been committed.
Topology First method provides a relational topology $T_i$ together with a soft spatial scaffold $\bar{G}_i$, whereas Text First method provides a browser-validated geometry $G_i$ that should remain fixed. 
The Commit \& Assemble Phase therefore realizes visual appearance under these different levels of geometric authority: the former allows metric adaptation in exchange for stronger scene integration, while the latter preserves exact registration at the cost of weaker cross-layer integration.

\noindent \paragraph{\textbf{Joint-canvas synthesis for Topology First.}}
For Topology First, the Visual Synthesis Agent jointly realizes the background and visual regions on a shared canvas. 
Given region-level visual briefs $U_i$, the canonical background $B$, and the topology-derived spatial scaffold, we generate
\begin{equation}
A_i^{\mathrm{topo}}
=
D_{\mathrm{canvas}}
!\left(
B,V,U_i,m(T_i,\bar{G}_i)
\right),
\end{equation}
where $m(T_i,\bar{G}_i)$ provides soft spatial control derived from the grounded composition. 
Because foreground elements and their surrounding background are generated together, texture, illumination, decoration, and spatial emphasis can be coordinated at the scene level, producing stronger visual integration.

The same joint generation, however, does not guarantee exact metric registration.
Generated elements may shift or reshape relative to $\bar{G}_i$, as the diffusion process does not always strictly adhere to the guiding masks or coordinate-integrated prompts. 
We therefore treat $\bar{G}_i$ as a scaffold rather than a geometric contract and align native text and holders to the realized visual regions after synthesis. 
Importantly, this adjustment refines metric placement without changing the committed relational topology $T_i$.

\noindent \paragraph{\textbf{Geometry-registered synthesis for Text First.}}
Text First enters this phase with a browser-validated geometric contract $G_i$. 
Rather than allowing image generation to redefine this geometry, the Visual Synthesis Agent converts the registered visual regions into a spatial generation mask $\Omega_i=m(G_i)$ on a shared canvas. 
The masked regions are then synthesized jointly under their corresponding visual briefs:
\begin{equation}
\widetilde{A}_i
=
D_{\mathrm{mask}}
\left(
 B,V,U_i,\Omega_i
\right),
\end{equation}
where the mask fixes the spatial support assigned to each visual region while diffusion determines its visual realization.
Considering the inherently shift-prone nature of the diffusion process, we apply the deterministic, slot-aware matting procedure (detailed in Appendix~\ref{app:matting}) to extract generated content within each registered region. 
The extraction step is what turns a pixel-space generation into a geometry-conforming, independently placeable asset.

After generation, the system extracts the synthesized content within each registered region into a separate transparent layer,
\begin{equation}
a_{ij}
=
\operatorname{extract}
\left(
\widetilde{A}_i,g_{ij}
\right),
\qquad j\in\mathcal{V}_i,
\end{equation}
and places these layers back at the geometry specified by $G_i$. 
This preserves the spatial organization validated by Composition-Inspired Grounding Agent while turning each generated visual into an independently selectable and editable slide object. 
Consequently, this geometry-registered construction provides stronger metric control and object-level editability.

\noindent \paragraph{\textbf{Constrained assembly and refinement.}}
Both routes finally assemble generated visuals with editable native text into the slide representation. 
Text, its associated holder, and local decoration are treated as a single content unit so that subsequent refinement cannot break their spatial correspondence. 
Deterministic fitting is applied first to resolve local sizing and overflow, after which the Judge Agent may propose only localized modifications that respect the structure already committed by the corresponding route. 
Each candidate modification is re-rendered and accepted only when it improves the slide without introducing collision, semantic loss, or violations of the committed topology or geometry.

The judge therefore acts as a constrained refiner rather than a second layout planner. 
Topology First method preserves the discovered relational organization while allowing limited geometric adaptation to the realized canvas; Text First method preserves the HTML-derived geometric contract throughout synthesis and assembly. 
Together, the two strategies expose complementary operating points between scene-level visual integration and precise, editable geometric control.

\begin{table*}[!t]
    \centering
    \caption{
System-level results on our test cases mentioned in~\ref{sec:exp-setup}. 
Higher is better for all quality metrics; lower is better for cost and latency. 
We mark the winner and second winner within each Language Model group for each quality metric using \winnerlegend and \secondwinnerlegend, respectively.
For each quality-metric column, the best and second-best values are shown in \textbf{\emph{bold}} and \underline{underlined}, respectively.}
    \label{tab:system-results}
    \footnotesize
    \setlength{\tabcolsep}{2.3pt}
    \renewcommand{\arraystretch}{0.98}
    \setlength{\aboverulesep}{0.30ex}
    \setlength{\belowrulesep}{0.30ex}
    \setlength{\cmidrulesep}{0.20ex}
    \resizebox{\textwidth}{!}{%
    \begin{tabular}{@{}
        >{\centering\arraybackslash}m{2.55cm}
        >{\centering\arraybackslash}m{2.3cm}
        >{\centering\arraybackslash}m{2.1cm}
        *{5}{>{\centering\arraybackslash}m{1.1cm}}
        >{\centering\arraybackslash}m{1.4cm}
        >{\centering\arraybackslash}m{1.6cm}
    @{}}
        \toprule
        \multicolumn{3}{c}{\textbf{Configuration}} & \multicolumn{5}{c}{\textbf{Quality Metrics}} & \multicolumn{2}{c}{\textbf{Efficiency}} \\[0.35ex]
        \cmidrule(lr){1-3} \cmidrule(lr){4-8} \cmidrule(lr){9-10}
        \textbf{System} & \textbf{Language Model} & \textbf{Image Model} & \textbf{VE $\uparrow$} & \textbf{IF $\uparrow$} & \textbf{Content $\uparrow$} & \textbf{Layout $\uparrow$} & \textbf{Topo-Vendi $\uparrow$} & \textbf{Cost (USD) $\downarrow$} & \textbf{Latency (s) $\downarrow$} \\[0.20ex]
        \midrule
        \multirow{3}{*}{\textbf{\textsc{AutoPresent}}} & GPT-5.4 & -- & 2.21 & 3.35 & 3.38 & 0.69 & 0.79 & \multirow{3}{*}{0.12} & \multirow{3}{*}{329.64} \\
        & Gemini 3.1 Pro & -- & 2.79 & 4.21 & 3.56 & 0.93 & 0.42 & & \\
        & Qwen 3.8 Max & -- & 2.56 & 3.28 & 3.07 & 0.70 & 0.60 & & \\
        \midrule
        \multirow{3}{*}{\textbf{\textsc{DeepSlides}}} & GPT-5.4 & -- & 2.00 & 3.06 & 2.57 & 0.55 & 0.66 & \multirow{3}{*}{1.21} & \multirow{3}{*}{1006.75} \\
        & Gemini 3.1 Pro & -- & 2.86 & 3.65 & 3.04 & \secondwinner{\underline{0.98}} & 0.71 & & \\
        & Qwen 3.8 Max & -- & 2.17 & 3.11 & 2.58 & 0.84 & 0.68 & & \\
        \midrule
        \multirow{6}{*}{\shortstack[c]{\textbf{\textsc{DeepPresenter}}\\[-1pt](With Image Generation)}}
        & \multirow{2}{*}{GPT-5.4} & GPT Image 2 & 3.06 & \secondwinner{4.76} & 3.89 & 0.83 & 0.58 & \multirow{6}{*}{1.02} & \multirow{6}{*}{1317.50} \\
        & & Nano Banana Pro & 2.83 & 4.53 & 3.67 & 0.77 & 0.61 & & \\
        & \multirow{2}{*}{Gemini 3.1 Pro} & GPT Image 2 & 2.88 & 4.19 & 3.28 & 0.77 & 0.59 & & \\
        & & Nano Banana Pro & 2.97 & 4.32 & 3.40 & 0.76 & 0.58 & & \\
        & \multirow{2}{*}{Qwen 3.8 Max} & GPT Image 2 & 2.42 & 3.35 & 2.76 & 0.60 & 0.49 & & \\
        & & Nano Banana Pro & 2.38 & 3.43 & 2.79 & 0.61 & 0.51 & & \\
        \midrule
        \multirow{6}{*}{\shortstack[c]{\textbf{\textsc{SlidesWeaver}}\\[-1pt](Topology First)}}
        & \multirow{2}{*}{GPT-5.4} & GPT Image 2 & \winner{\underline{3.42}} & 4.50 & 3.60 & 0.84 & 0.85 & \multirow{6}{*}{0.64} & \multirow{6}{*}{556.80} \\
        & & Nano Banana Pro & 2.71 & 4.04 & 3.42 & 0.55 & \winner{0.85} & & \\
        & \multirow{2}{*}{Gemini 3.1 Pro} & GPT Image 2 & 2.92 & 3.89 & 3.01 & 0.90 & 0.84 & & \\
        & & Nano Banana Pro & 2.69 & 3.83 & 3.04 & 0.75 & \secondwinner{0.85} & & \\
        & \multirow{2}{*}{Qwen 3.8 Max} & GPT Image 2 & \secondwinner{3.12} & 4.38 & 3.39 & 0.75 & 0.85 & & \\
        & & Nano Banana Pro & 2.60 & 3.85 & 3.17 & 0.54 & \secondwinner{0.83} & & \\
        \midrule
        \multirow{6}{*}{\shortstack[c]{\textbf{\textsc{SlidesWeaver}}\\[-1pt](Text First)}}
        & \multirow{2}{*}{GPT-5.4} & GPT Image 2 & \secondwinner{3.29} & \winner{\textbf{4.92}} & \winner{\textbf{4.28}} & \winner{0.98} & \secondwinner{0.84} & \multirow{6}{*}{0.88} & \multirow{6}{*}{1279.86} \\
        & & Nano Banana Pro & 2.96 & 4.71 & \secondwinner{\underline{4.04}} & \secondwinner{0.89} & 0.83 & & \\
        & \multirow{2}{*}{Gemini 3.1 Pro} & GPT Image 2 & \winner{3.32} & \winner{4.79} & \secondwinner{3.79} & \winner{\textbf{0.99}} & 0.84 & & \\
        & & Nano Banana Pro & \secondwinner{2.99} & \secondwinner{4.69} & \winner{3.81} & 0.97 & \winner{\textbf{0.87}} & & \\
        & \multirow{2}{*}{Qwen 3.8 Max} & GPT Image 2 & \winner{\textbf{3.44}} & \winner{\underline{4.88}} & \winner{3.93} & \winner{0.97} & \winner{\underline{0.85}} & & \\
        & & Nano Banana Pro & 2.89 & \secondwinner{4.50} & \secondwinner{3.71} & \secondwinner{0.90} & 0.81 & & \\
        \bottomrule
    \end{tabular}%
    }
\end{table*}

\section{Experiments}
\label{sec:experiments}

We evaluate \textsc{SlidesWeaver} along three complementary dimensions: overall presentation quality, the contribution of diffusion-based composition planning, and reliability and practical efficiency.
Specifically, we investigate the following research questions.

\noindent\textbf{RQ1: Overall presentation quality.}
How do \textsc{SlidesWeaver} and existing presentation generators compare in visual expressiveness, instruction following, content quality, and layout validity?

\noindent\textbf{RQ2: Diffusion-based composition planning.}
Does diffusion-based composition planning produce slide topologies that are more diverse while remaining layout-valid?

\noindent\textbf{RQ3: Reliability and practical efficiency.}
Does \textsc{SlidesWeaver} achieve greater visual expressiveness without compromising slide validity or practical efficiency, as measured by severe layout defects, first-pass success rate, generation latency, and monetary cost?

\subsection{Experiment Setup}
\label{sec:exp-setup}

\noindent \paragraph{\textbf{Models and Baselines.}}
We compare \textsc{SlidesWeaver} against three academic presentation generation systems: \textsc{AutoPresent}, which constructs structured visual programs~\cite{ge2025autopresent}; \textsc{DeepSlides}, which separates slide design from code realization~\cite{cui2026deepslides}; and \textsc{DeepPresenter}, which adopts an environment-grounded researcher-presenter pipeline~\cite{zheng2026deeppresenter}.
For all methods, we evaluate three text backbones: \texttt{GPT 5.4}, \texttt{Gemini 3.1 Pro}, and \texttt{Qwen 3.8 Max}. 
For the image-enabled \textsc{DeepPresenter} and \textsc{SlidesWeaver} (Text First and Topology First), each text backbone is paired with either \texttt{GPT Image 2} or \texttt{Nano Banana Pro}. 
This results in 3 configurations for each image-free baseline and 6 configurations for each image-enabled system, for 24 configurations in total.
Since all evaluated systems produce editable PPTX files, editability is treated as an output requirement rather than a comparative quality dimension.

\noindent \paragraph{\textbf{Test Set.}}
Following evaluation settings in~\cite{zheng2025pptagent,zheng2026deeppresenter}, we construct a test set of 72 presentation requests covering six domains including Science, Technology, Business, Education, Society, and Culture / Creative. 
The test set contains two complementary tracks. 
The \emph{\textbf{topic-driven}} track asks systems to develop a presentation from a given topic, while the \emph{\textbf{goal-oriented}} track additionally specifies an explicit communication objective. 
Across both tracks, we consider six common presentation goals: explaining a concept, teaching an audience, persuading, comparing alternatives or supporting a decision, reporting evidence, and narrating or inspiring. 
The benchmark is balanced across the two tracks and six domains, with 36 cases per track and 12 cases per domain. 
The cases also vary in compositional difficulty, including 17 low-, 29 medium-, and 26 high-difficulty requests. 
Each case specifies the topic, target audience, communication goal, success criterion, style brief, language, and requested number of slides.

\noindent \paragraph{\textbf{Evaluation Protocol.}}
We evaluate all 24 model-system configurations on the full 72-case test set, resulting in 1,728 generation cases. 
We record the generated artifacts, generation events, latency, and failures for every run. 
For quality evaluation, a common blinded judge based on \texttt{GPT 5.6 Terra} receives the original case brief, extracted slide text, artifact-level evidence, and the ordered slide montage. 
Following established presentation-generation evaluation protocols and recent benchmarks \cite{ge2025autopresent, zheng2025pptagent, cui2026deepslides, yang2026slidesgen, chen2026presentbench, chen2026xslides, jang2026deckbench}, we evaluate each generated deck along the following dimensions:
\begin{itemize}[leftmargin=*]
\item \textbf{Visual Expressiveness (VE).} Measures the slide design's visual richness, composition quality, and effectiveness on a 1-5 scale.
\item \textbf{Instruction Following (IF).}
Measures how well the generated deck follows the requested topic, audience, communication goal, style brief, and slide-count constraints on a 1-5 scale.
\item \textbf{Content Quality.}
Measures the correctness, relevance, organization, and narrative coherence of the presentation content on a 1-5 scale~\cite{ge2025autopresent,zheng2025pptagent,cui2026deepslides}.
\item \textbf{Layout Validity.}
For each successfully generated and rendered slide, we detect severe overlap, clipping, and out-of-bounds elements.
Let $\mathcal{V}_{ki}$ denote this set of slides; the deck-level score is the fraction free from all three defects:
\begin{equation}
    \label{eq:layout}
    \mathrm{Layout}_{ki}
    =
    \frac{1}{|\mathcal{V}_{ki}|}
    \sum_{j\in\mathcal{V}_{ki}}
    \mathbf{1}\!\left[
        \neg O_{kij}
        \wedge \neg C_{kij}
        \wedge \neg B_{kij}
    \right],
    \qquad |\mathcal{V}_{ki}| > 0.
\end{equation}

\item \textbf{Normalized Topology Vendi (Topo-Vendi).}
We measure within-deck structural diversity by representing eligible content slides as topology graphs encoding region types, size classes, groupings, and directed spatial relations.
Using two iterations of Weisfeiler--Lehman graph similarity, we compute the Vendi Score $\mathrm{VS}$~\cite{friedman2022vendiscore} and report its normalized log form $\log(\mathrm{VS})/\log |E_{ki}|\in[0,1]$:
\begin{equation}
    \mathrm{TopoVendi}_{ki}
    =
    \frac{\log \mathrm{VS}_{ki}}{\log |E_{ki}|}
    =
    \frac{
    -\sum_r \lambda_r \log \lambda_r
    }{
    \log |E_{ki}|
    },
\end{equation}
where $\{\lambda_r\}$ are the normalized eigenvalues of the similarity matrix and $E_{ki}$ is the set of eligible content slides.

\item \textbf{Cost and Latency.}
We record the total API cost of all LLM, VLM, and image-generation calls and the end-to-end wall-clock generation time for each deck for later comparison.
\end{itemize}
For each configuration, we report means over the 72 test cases.

\subsection{Results and Discussion}

Table~\ref{tab:system-results} summarizes the results over all 72 presentation requests. 
Each row gives a model configuration. For general comparison, we average cost and latency over each system's configurations.

\noindent \paragraph{\textbf{RQ1: Overall presentation quality.}}
\textsc{SlidesWeaver} shows a clear overall quality advantage.
Across the 15 language-model--metric comparisons in Table~\ref{tab:system-results}, every best score is achieved by one of our two routes.
The two routes represent complementary generation strategies: Text First emphasizes faithful realization of pre-committed text and geometry, whereas Topology First allows the visual composition to play a stronger role in shaping the final slide.
For the text-preservation-oriented Text First route, averaging over its six configurations yields improvements of 14.2\% in VE, 15.9\% in IF, 17.7\% in Content, 20.3\% in Layout, and 22.9\% in Topo-Vendi over the strongest baseline for each metric.
Together, the two routes enable \textsc{SlidesWeaver} to combine strong content quality and instruction following with expressive and structurally diverse slide designs, rather than excelling in only a single quality dimension.

The metric-level results further reveal how these gains arise.
Visual expressiveness is naturally sensitive to the underlying image backbone.
Within \textsc{SlidesWeaver}, GPT Image 2 achieves higher VE than Nano Banana Pro in every matched setting.
Importantly, our pipeline consistently translates this stronger image-generation capability into higher presentation-level visual expressiveness, suggesting that \textsc{SlidesWeaver} can effectively exploit improvements in the underlying image model rather than bottlenecking them during slide construction.
For instruction following, the advantage is more structural. 
The generation process is decomposed into small, well-specified stages, with each agent responsible for a focused subtask and explicit topology or geometry providing additional constraints.
The two routes then make different commitments within this shared framework.
Text First fixes generated text and geometry before visual synthesis, naturally favoring text preservation and stable layout, while Topology First preserves greater freedom for composition-driven generation.
The latter thus depends more on the image model's spatial controllability; for example, Nano Banana Pro more readily deviates from the prescribed topology, explaining its lower Layout scores in several Topology First settings.

\noindent \paragraph{\textbf{RQ2: Topology and layout diversity.}}
Diffusion-based composition planning substantially expands the structural design space.
Topology First and Text First achieve average Topo-Vendi scores of 0.845 and 0.840, respectively, compared with 0.683 for the strongest baseline, corresponding to gains of 23.7\% and 22.9\%.
Importantly, this improvement reflects \emph{topological} rather than merely stylistic diversity: the generated slides vary in grouping, hierarchy, and spatial organization, rather than only in colors or decorative assets.
The nearly identical Topo-Vendi scores of the two routes are particularly informative, as they show that the diversity persists under different downstream realization strategies when translating diffusion proposals into precise geometry.
This suggests that the shared diffusion proposal itself provides the structural diversity prior, rather than the effect arising from a particular commitment mechanism.
Moreover, Text First retains a Topo-Vendi of 0.840 while achieving an average Layout score of 0.950, demonstrating that the expanded design space can remain compatible with highly valid slide layouts.
We further isolate the contribution of diffusion-based composition planning in the ablation study.

\noindent \paragraph{\textbf{RQ3: Reliability and practical efficiency.}}
\begin{figure}[!ht]
    \vspace{-5pt}
    \centering
    \includegraphics[width=0.75\columnwidth]{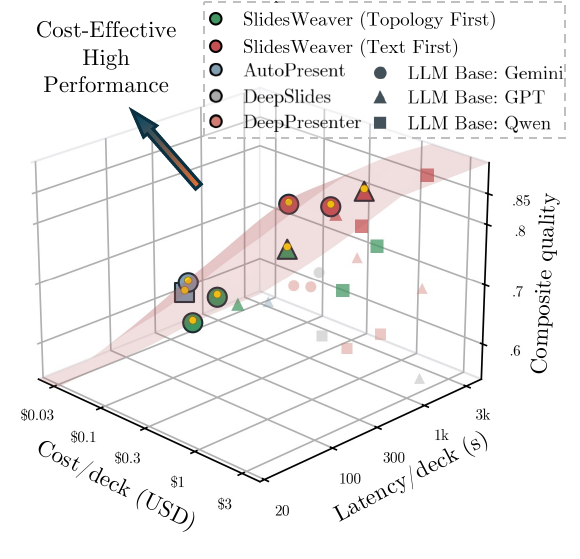}
    \caption{
\textsc{SlidesWeaver} dominates Pareto-optimal configurations; the other two are low-cost AutoPresent variants that sacrifice performance.
We plot all 24 system-model configurations by cost, latency, and normalized composite quality; dark outlines mark Pareto-optimal points.
}
    \label{fig:efficiency-pareto}
    \vspace{-5pt}
\end{figure}
The quality gains of \textsc{SlidesWeaver} are achieved with favorable end-to-end efficiency. 
Topology First costs \$0.64 per deck and takes 556.8 seconds, making it 37.3\% cheaper and 57.7\% faster than DeepPresenter, while Text First delivers the strongest overall quality at \$0.88 per deck and 1,279.9 seconds, remaining 13.7\% cheaper and 2.9\% faster. 
Consequently, six of the eight non-dominated configurations in Figure~\ref{fig:efficiency-pareto} belong to \textsc{SlidesWeaver}, pushing the quality-cost-latency frontier beyond existing systems. 
This is particularly notable because \textsc{SlidesWeaver} performs more involved visual synthesis than DeepPresenter, which incorporates images through text-only image tool calls.
Our pipeline additionally requires mask-conditioned generation and subsequent asset extraction to preserve planned geometry and editability. 
Despite additional input tokens, \textsc{SlidesWeaver} remains cheaper and faster end-to-end. 
The system is also highly reliable, with first-pass success rates of 97.69\% for Topology First and 96.99\% for Text First, demonstrating that the proposed hybrid generation paradigm is efficient and robust in practical use.

\begin{figure*}[!ht]
    \centering
    \includegraphics[width=0.9\textwidth]{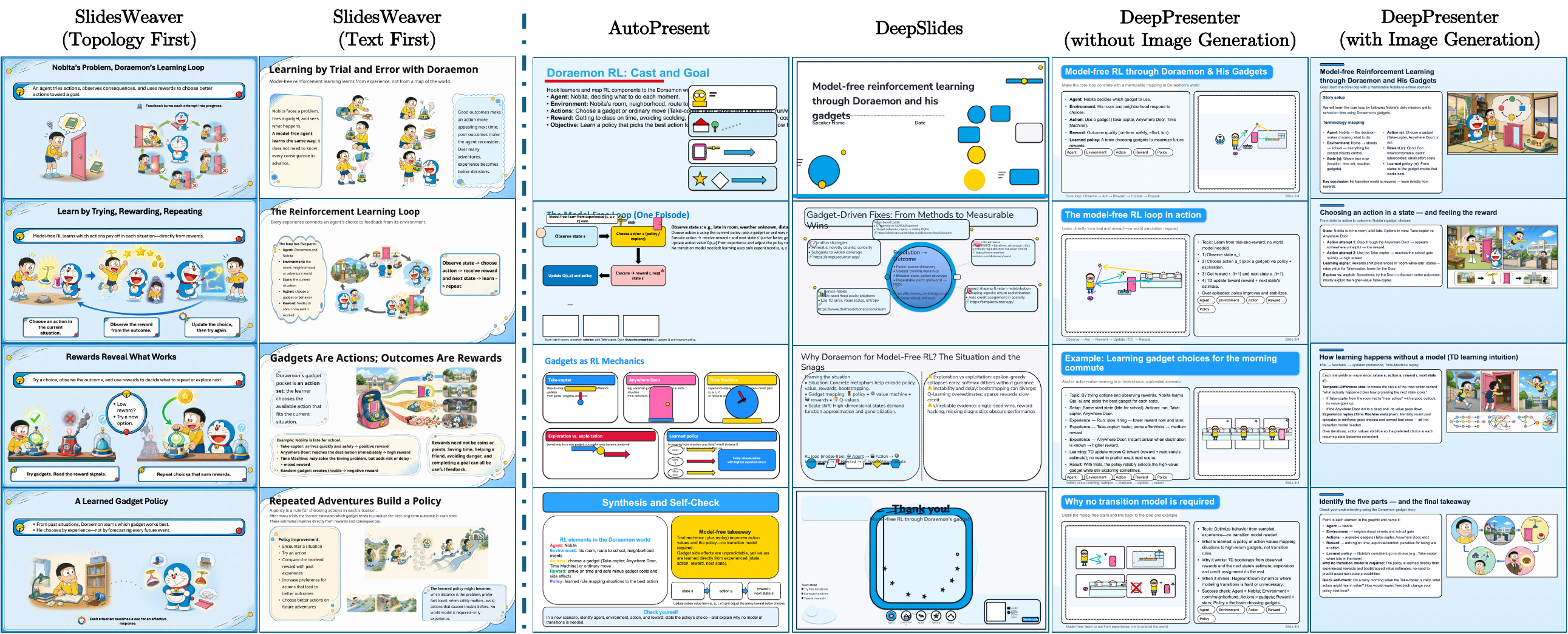}
    \caption{
    Qualitative comparison on a presentation request explaining model-free reinforcement learning through Doraemon and his gadgets.
    All systems use GPT-5.4 as the language model, and GPT Image 2 is used whenever image generation is applicable.
    From left to right, we show \textsc{SlidesWeaver} (Topology First), \textsc{SlidesWeaver} (Text First), AutoPresent, DeepSlides, DeepPresenter without/with image generation.
    While the baselines predominantly rely on recurring text-card or text-image partitions, \textsc{SlidesWeaver} exhibits greater variation in grouping, hierarchy, and spatial composition across slides.
    }
    \label{fig:case-study}
\end{figure*}

\subsection{Ablation Study}

\textsc{SlidesWeaver} is a tightly coupled pipeline in which several modules are indispensable for completing generation end-to-end.
We therefore ablate only components that admit meaningful alternatives, and focus on VE, Layout, and Topo-Vendi metrics, which are directly affected by visual planning and synthesis.

\begin{table}[!ht]
    \centering
    \vspace{-5pt}
    \caption{
    Ablation of diffusion-based composition planning and the layout judge in Text First.
    All upstream outputs are reused.
For w/o Diffusion, HTML planning uses the same prompt without the diffusion-generated skeleton; Topology First is excluded as its planning is intrinsically diffusion-based.
    \ding{168}/\ding{170}: GPT Image 2/Nano Banana Pro; O, w/o D, and w/o J: original, without diffusion, and without judge.
    }
    \label{tab:ablation-planning}
    \footnotesize
    \setlength{\tabcolsep}{0.2pt}
    \resizebox{\columnwidth}{!}{%
    \begin{tabular}{@{}
        >{\centering\arraybackslash}m{2.30cm}|
        *{3}{>{\centering\arraybackslash}m{0.72cm}}|
        *{3}{>{\centering\arraybackslash}m{0.72cm}}|
        *{3}{>{\centering\arraybackslash}m{0.72cm}}
    @{} }
        \toprule
        & \multicolumn{3}{>{\centering\arraybackslash}m{2.16cm}|}{\textbf{VE $\uparrow$}} &
        \multicolumn{3}{>{\centering\arraybackslash}m{2.16cm}|}{\textbf{Layout $\uparrow$}} &
        \multicolumn{3}{>{\centering\arraybackslash}m{2.16cm}}{\textbf{Topo-Vendi $\uparrow$}} \\
        \cmidrule(lr){2-4} \cmidrule(lr){5-7} \cmidrule(lr){8-10}
        \textbf{System}
        & \textbf{O} & \textbf{w/o D} & \textbf{w/o J}
        & \textbf{O} & \textbf{w/o D} & \textbf{w/o J}
        & \textbf{O} & \textbf{w/o D} & \textbf{w/o J} \\
        \midrule
        GPT-5.4\textsuperscript{\scriptsize\ding{168}}
        & 3.29 & 2.99 & 3.14 & 0.98 & 0.97 & 0.92 & 0.84 & 0.54 & 0.85 \\
        GPT-5.4\textsuperscript{\scriptsize\ding{170}}
        & 2.96 & 2.61 & 2.81 & 0.89 & 0.88 & 0.83 & 0.83 & 0.53 & 0.83 \\
        Gemini 3.1 Pro\textsuperscript{\scriptsize\ding{168}}
        & 3.32 & 2.98 & 3.17 & 0.99 & 0.98 & 0.87 & 0.84 & 0.54 & 0.84 \\
        Gemini 3.1 Pro\textsuperscript{\scriptsize\ding{170}}
        & 2.99 & 2.65 & 2.84 & 0.97 & 0.96 & 0.85 & 0.87 & 0.57 & 0.87 \\
        Qwen 3.8 Max\textsuperscript{\scriptsize\ding{168}}
        & 3.44 & 3.11 & 3.29 & 0.97 & 0.96 & 0.85 & 0.85 & 0.55 & 0.85 \\
        Qwen 3.8 Max\textsuperscript{\scriptsize\ding{170}}
        & 2.89 & 2.63 & 2.84 & 0.90 & 0.89 & 0.78 & 0.81 & 0.51 & 0.82 \\
        \midrule
        \textbf{Average}
        & \textbf{3.15} & 2.83 & 3.02
        & \textbf{0.95} & 0.94 & 0.85
        & \textbf{0.84} & 0.54 & 0.84 \\
        \bottomrule
    \end{tabular}}
    \vspace{-5pt}
\end{table}

Table~\ref{tab:ablation-planning} isolates the roles of diffusion and judging.
Removing diffusion reduces Topo-Vendi from 0.84 to 0.54 and VE from 3.15 to 2.83, while Layout remains almost unchanged (0.95 vs.\ 0.94), directly confirming that the diffusion proposal is the primary source of compositional diversity.
Removing the judge shows the complementary effect: Topo-Vendi remains essentially unchanged, whereas Layout drops from 0.95 to 0.85.
The degradation is smallest with GPT-5.4 ($-0.06$ on average) and larger with Gemini and Qwen ($-0.12$), suggesting the judge mainly offsets differences in one-shot HTML planning capability rather than creating diversity itself.

\begin{table}[t]
    \centering
    \caption{
    Ablation of the final synthesis path.
    All outputs before visual synthesis are reused; only the realization strategy changes.
    \ding{46} and \ding{111} denote Inpainting and Layered Generation, respectively; * marks the original choice.
    }
    \label{tab:ablation-synthesis}
    \footnotesize
    \setlength{\tabcolsep}{0.2pt}
    \resizebox{\columnwidth}{!}{%
    \begin{tabular}{@{}
        >{\centering\arraybackslash}m{1.85cm}|
        *{2}{>{\centering\arraybackslash}m{0.58cm}}|
        *{2}{>{\centering\arraybackslash}m{0.58cm}}|
        *{2}{>{\centering\arraybackslash}m{0.58cm}}|
        *{2}{>{\centering\arraybackslash}m{0.58cm}}
    @{}}
        \toprule
        & \multicolumn{4}{>{\centering\arraybackslash}m{2.32cm}|}{\textbf{Topo First}} &
        \multicolumn{4}{>{\centering\arraybackslash}m{2.32cm}}{\textbf{Text First}} \\
        \cmidrule(lr){2-5} \cmidrule(lr){6-9}
        \textbf{System}
        & \multicolumn{2}{>{\centering\arraybackslash}m{1.16cm}|}{\textbf{VE $\uparrow$}} &
        \multicolumn{2}{>{\centering\arraybackslash}m{1.16cm}|}{\textbf{Layout $\uparrow$}} &
        \multicolumn{2}{>{\centering\arraybackslash}m{1.16cm}|}{\textbf{VE $\uparrow$}} &
        \multicolumn{2}{>{\centering\arraybackslash}m{1.16cm}}{\textbf{Layout $\uparrow$}} \\
        \cmidrule(lr){2-3} \cmidrule(lr){4-5} \cmidrule(lr){6-7} \cmidrule(lr){8-9}
        & \textsuperscript{\scriptsize\ding{46}*} & \textsuperscript{\scriptsize\ding{111}}
        & \textsuperscript{\scriptsize\ding{46}*} & \textsuperscript{\scriptsize\ding{111}}
        & \textsuperscript{\scriptsize\ding{111}*} & \textsuperscript{\scriptsize\ding{46}}
        & \textsuperscript{\scriptsize\ding{111}*} & \textsuperscript{\scriptsize\ding{46}} \\
        \midrule
        GPT-5.4\textsuperscript{\scriptsize\ding{168}}
        & 3.42 & 3.39 & 0.84 & 0.90 & 3.29 & 3.34 & 0.98 & 0.90 \\
        GPT-5.4\textsuperscript{\scriptsize\ding{170}}
        & 2.71 & 3.02 & 0.55 & 0.82 & 2.96 & 2.66 & 0.89 & 0.65 \\
        Gemini 3.1 Pro\textsuperscript{\scriptsize\ding{168}}
        & 2.92 & 3.27 & 0.90 & 0.94 & 3.32 & 3.34 & 0.99 & 0.89 \\
        Gemini 3.1 Pro\textsuperscript{\scriptsize\ding{170}}
        & 2.69 & 2.87 & 0.75 & 0.87 & 2.99 & 3.14 & 0.97 & 0.84 \\
        Qwen 3.8 Max\textsuperscript{\scriptsize\ding{168}}
        & 3.12 & 3.31 & 0.75 & 0.86 & 3.44 & 3.05 & 0.97 & 0.81 \\
        Qwen 3.8 Max\textsuperscript{\scriptsize\ding{170}}
        & 2.60 & 2.89 & 0.54 & 0.83 & 2.89 & 2.55 & 0.90 & 0.67 \\
        \midrule
        \textbf{Average}
        & 2.91 & 3.13 & 0.72 & 0.87
        & 3.15 & 3.01 & 0.95 & 0.79 \\
        \bottomrule
    \end{tabular}}
    \vspace{-10pt}
\end{table}

Table~\ref{tab:ablation-synthesis} explores whether alternative downstream synthesis strategies can further improve the two generation routes.
For Topology First, layered generation improves Layout in all six configurations (0.72 to 0.87 on average), while VE generally increases but remains model-dependent; when inpainting already integrates foreground and background well, explicit layering can instead slightly reduce visual coherence.
Conversely, replacing layered generation with inpainting in Text First consistently reduces Layout (0.95 to 0.79), while VE varies considerably across model combinations and decreases from 3.15 to 3.01 on average.
This indicates that inpainting quality depends strongly on the image model's spatial and prompt-following capability: once registered geometry is handed back to image synthesis, subsequent visual drift leaves limited room for the planning-stage judge to correct it.

\subsection{Qualitative Case Study}

Figure~\ref{fig:case-study} highlights the distinction between \emph{using generated images} and \emph{using generation for visual composition}.
While DeepPresenter constantly adds semantically relevant illustrations to almost predefined text-image regions, \textsc{SlidesWeaver} uses generated visual structure to jointly organize text, illustrations, grouping, and spatial layout.
Across both generation routes, \textsc{SlidesWeaver} yields richer and more diverse yet coherent slide structures.
The case study thus supports RQ1 and RQ2: diffusion provides richer visual expressions and a composition prior for presentation design.
\section{Conclusion}

We presented \textsc{SlidesWeaver}, a hybrid multi-agent system that combines diffusion-based pixel-space exploration with structured presentation slides construction through staged multimodal commitment. 
Across 72 diverse cases, its two generation routes deliver leading visual, semantic, layout, and topological quality while preserving editable text and geometry.
The slide-separable design further supports practical parallel execution with favorable cost and latency, providing a strong foundation for efficient, visually rich, and editable presentation generation.


\section*{Ethical Considerations}
\textsc{SlidesWeaver} is intended as an assistive presentation-authoring system rather than a replacement for human verification and editorial judgment. Like other systems built on large language and image-generation models, it may produce inaccurate, biased, or inappropriate textual and visual content~\cite{zhou2026flattery}, and generated assets may inadvertently resemble copyrighted, trademarked, or otherwise protected material. Users should therefore review generated presentations for factual correctness, attribution, licensing, and suitability before use or distribution. The stylized examples in this work are used only to study presentation-generation capabilities and should not be interpreted as granting rights to reproduce protected characters, brands, or artistic styles. More broadly, editability is intended to aid such human review and correction before deployment.

\newpage
\bibliographystyle{ACM-Reference-Format}
\bibliography{sample-base}

\newpage
\appendix

\section{Scope of Hybrid Generation: Visual-Tool Use in \textsc{DeepPresenter}}
\label{app:deeppresenter-visual-tools}

\textsc{DeepPresenter} provides its Research Agent with both image retrieval and image-generation tools before the Presenter Agent constructs slides programmatically in HTML~\cite{zheng2026deeppresenter}.
The availability of both tool types, however, does not by itself constitute pipeline-level hybrid generation.
In this work, we use ``hybrid'' specifically to denote the integration of visual generation into composition planning and structured slide realization, rather than its optional use as a source of raster assets.

To characterize how visual generation is used under this distinction, we probe \textsc{DeepPresenter} on ten randomly selected evaluation cases with three consistent language-model backbones in Table~\ref{tab:system-results}, exposing both image search and generation in every run.
We use local stubs that preserve tool interfaces and record action traces without invoking remote visual APIs, thereby avoiding differences in retrieved or generated artifact quality from influencing tool-selection behavior.
Across all 30 runs, only two action patterns occur:
\texttt{search $\rightarrow$ download $\rightarrow$ generation} in 23 runs and
\texttt{search $\rightarrow$ generation} in the remaining 7.
No run invokes generation before search or interleaves search and generation, and all 117 image-generation events occur after the final \texttt{search\_images} action of the corresponding job.
The observed policy therefore consistently follows a \emph{search-first, generation-later} sequence, treating generation as a subsequent visual-asset acquisition step rather than an initial mechanism for exploring slide composition.
In our evaluation presented in Table~\ref{tab:system-results}, we disable the image-search API to remove retrieval as a competing visual source, thereby giving \textsc{DeepPresenter} maximal opportunity to exploit image generation and enabling a fairer comparison of visual-generation capability.

This behavior is consistent with the architectural distinction underlying our hybrid-generation claim.
In \textsc{DeepPresenter}, searched and generated images are collected as visual assets, while the downstream programmatic pipeline determines slide composition and geometry.
In \textsc{SlidesWeaver}, visual generation instead participates on both sides of structural commitment: before commitment, diffusion explores composition and topology in pixel space; after commitment, it synthesizes visually rich and instructive content under the established semantic and geometric structure.
Our contribution therefore lies in integrating visual generation throughout composition planning, structured synthesis, and editable assembly, rather than merely exposing image generation as an optional tool within a programmatic presentation pipeline.

\section{Slot-aware Chroma Matting}
\label{app:matting}

The Topology First and the Text First use the same slot-aware extraction pipeline.
The generated asset board is intended to have a uniform key-color field outside the registered slots, but diffusion can introduce color drift, illumination variation, anti-aliased boundaries, and chroma spill.
The extractor therefore estimates the realized key color from the generated board instead of comparing pixels with a fixed RGB constant.

Let $C\in[0,1]^{H\times W\times3}$ be the generated board and let $\partial C$ be a narrow border band.
The implementation uses a 12-pixel border band, clamped for small images.
We estimate the key color $K$ as the component-wise median of $\partial C$ and compute the Lab distance
\begin{equation}
    d_p=\left\|\operatorname{Lab}(C_p)-\operatorname{Lab}(K)\right\|_2 .
\end{equation}
The border-distance distribution gives image-specific thresholds for separating the key field from foreground content.
The implementation derives them as
\begin{equation}
\begin{aligned}
\tau_s&=\operatorname{clip}(P_{95}(d_{\partial C})+2,4,12),\\
\tau_g&=\max\{\tau_s+12,\min(35,P_{99.5}(d_{\partial C})+8)\},\\
\tau_f&=\max\{35,\tau_g\}.
\end{aligned}
\end{equation}
This staged threshold design keeps the coarse mask stable while reserving a narrower transition range for edge matting.

The coarse foreground mask is used only to recover topology.
Small connected components are discarded using both an absolute area floor and an image-area fraction, and components with very small width or height are ignored.
If one component overlaps multiple registered slots, we search a small set of increasingly strict distance thresholds for confident foreground cores.
We accept a split only when distinct cores have sufficient area and unique slot support.
The core search uses Lab-distance thresholds $50$, $70$, $90$, and $110$.
The ambiguous bridge is then partitioned by a nearest-core distance transform.
If these conditions are not met, the component remains intact instead of being split speculatively.

For each resulting component, we refine a local crop using a 5-pixel elliptical erosion and dilation band with 4 pixels of crop padding.
Interior pixels receive unit opacity, while edge opacity is the minimum of a Gaussian geometric alpha and a color-confidence alpha derived from $d_p$.
For partially transparent pixels, we estimate the foreground color with
\begin{equation}
    \widehat{F}_p=
    \frac{C_p-(1-\alpha_p)K}{\max(\alpha_p,\epsilon)} .
\end{equation}
We clip the result to the valid color range, apply a one-pixel alpha inset, zero very small alpha values, and remove residual key-color spill only on partially transparent pixels.
The earlier green-dominance variant and the current key-chroma projection implement the same decontamination role.
We use the projection form as the unified description because it also handles magenta or other selected key colors.

Finally, components are assigned back to slots.
Assignments first maximize component--slot overlap.
Unmatched components are greedily paired with missing slots by normalized centroid distance.
Matches beyond the distance gate are rejected.
Components assigned to one slot are alpha-composited and cropped to their nontransparent bounding box.
Holder layers are fitted to the exact target box, whereas visual layers are aspect-fitted inside the target box.
The manifest records the source component, crop, centroid, slot, and target box.
A slot without a reliable component is sent to bounded recovery, and the slide layout is never changed.

The procedure separates appearance recovery from geometric placement.
Geometry is used first to resolve ownership when generated components touch, and again to restore the planned slide coordinates after cropping.
This conservative policy never duplicates an ambiguously assigned component across slots, and it keeps recovery local to missing slots.

\begin{algorithm}[H]
\caption{Unified slot-aware chroma matting and placement}
\label{alg:slot-matting}
\begin{algorithmic}[1]
\Require generated board $C$, target boxes $G$, slot descriptors $U$
\State $K\gets\operatorname{MedianBorderColor}(C)$
\State $D\gets\operatorname{LabDistance}(C,K)$
\State $(\tau_s,\tau_g,\tau_f)\gets\operatorname{DualThresholds}(D_{\partial C})$
\State $Q\gets\operatorname{ConnectedComponents}(D\geq\tau_f)$
\State $Q\gets\operatorname{RemoveSmallComponents}(Q)$
\For{each component $q$ in $Q$}
    \If{$q$ overlaps multiple slots and distinct supported cores exist}
        \State $q\gets\operatorname{NearestCoreSplit}(q,U)$
    \EndIf
    \State $(\alpha,R)\gets\operatorname{EdgeMatte}(q,D,K,\tau_s,\tau_f)$
    \State $\widehat{C}_R\gets\operatorname{UnmixAndDespill}(C_R,K,\alpha)$
    \State emit $(\widehat{C}_R,\alpha,\operatorname{bbox}(q),\operatorname{centroid}(q))$
\EndFor
\State assign components to slots by overlap
\State greedily assign unmatched pairs using normalized centroid distance as the criterion
\For{each slot $u$ in $U$}
    \If{$u$ has no reliable assigned component}
        \State mark $u$ for bounded recovery
    \Else
        \State $L\gets\operatorname{AlphaCompositeAndCrop}(u)$
        \If{$u$ is a holder}
            \State $t\gets\operatorname{ExactBox}(u.target)$
        \Else
            \State $t\gets\operatorname{AspectFit}(L,u.target)$
        \EndIf
        \State emit $\operatorname{Place}(L,t,u.slot)$
    \EndIf
\EndFor
\State \Return transparent layers and missing-slot list
\end{algorithmic}
\end{algorithm}

\section{Local Diffusion for Composition Proposals}
\label{app:local-diffusion}

We use a lightweight local diffusion model, Z-Image-Turbo, to generate one 16:9 pixel-space composition proposal for each slide~\cite{imageteam2026zimageefficientimagegeneration}.
The prompt contains only the slide title, coarse communicative purpose, and concise visual logic, while detailed body text and exact coordinates are deliberately omitted to preserve compositional freedom.
The resulting raster is used only as a visual reference for grounding; readable text, executable geometry, and final visual appearance are determined by subsequent stages.

\begin{figure}[t]
    \centering
    \includegraphics[width=\columnwidth]{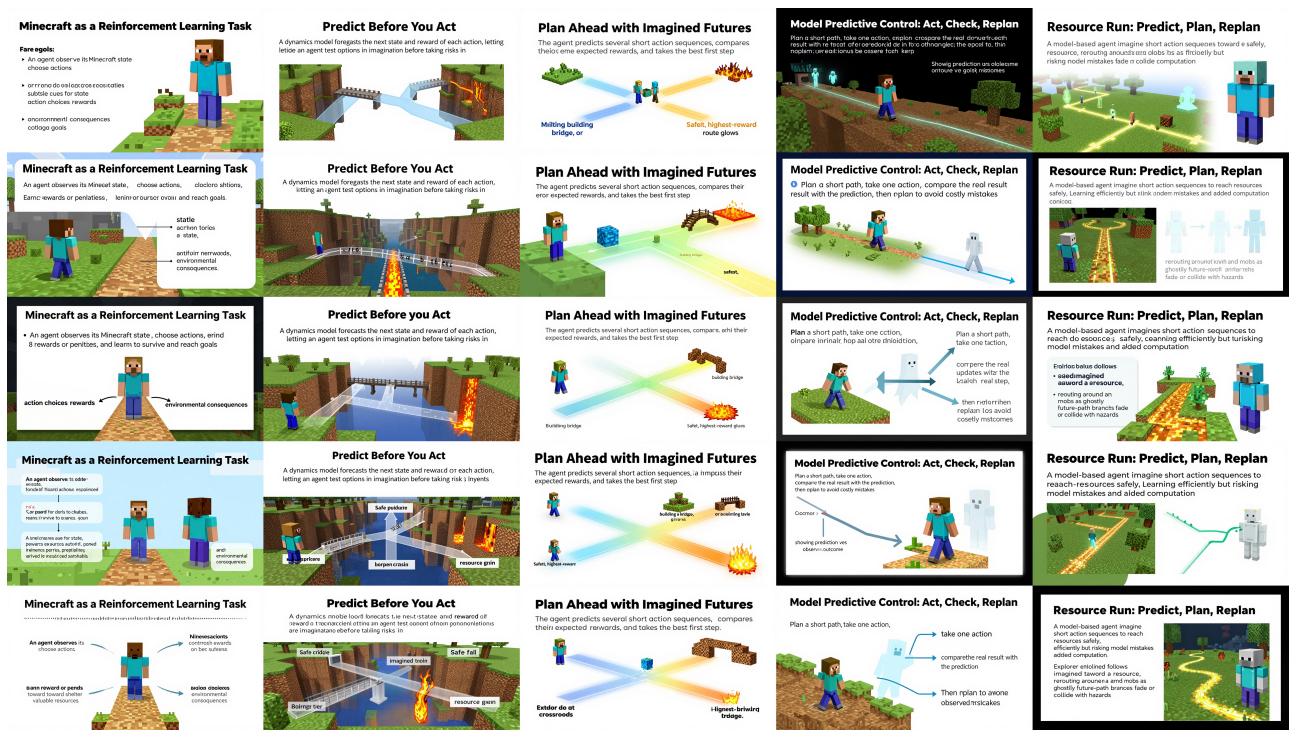}
    \caption{
    Composition proposals for five slide intents under different seeds.
    Each column fixes one slide intent, while rows vary only the seed: $\{43,48,53,58,63\}$.
    Local diffusion therefore explores substantially different groupings, hierarchies, and spatial organizations without changing the semantic input for a column.
    }
    \label{fig:local-diffusion-seeds}
\end{figure}

A key advantage of this proposal mechanism is its ability to explore diverse compositions with minimal intervention.
As illustrated in Figure~\ref{fig:local-diffusion-seeds}, changing only the random seed produces substantially different organizations for the same slide intent, varying grouping, hierarchy, spatial balance, asymmetry, and use of negative space.
This provides the grounding agents with a lightweight but expressive proposal space rather than forcing them to construct layouts directly from predefined geometric primitives.

The following block provides the complete Slide~01 prompt as our example for the composition-proposal stage.
\begin{tcolorbox}[
    breakable,
    colback=gray!5!white,
    colframe=black!75!black,
    title=Example Prompt for Slide 01,
    fonttitle=\bfseries,
    boxrule=1pt,
    arc=3mm,
    left=6pt,right=6pt,top=6pt,bottom=6pt
]
\small
Create a polished and visually creative 16:9 presentation slide.

Title: ``Minecraft as a Reinforcement Learning Task''

The slide communicates that An agent observes its Minecraft state, chooses actions, earns rewards or penalties, and learns to survive and reach goals.

Rough visual intent, offered as open-ended inspiration: ``A Minecraft-style adventurer facing a hazardous path toward shelter or valuable resources, with subtle cues for state, action choices, rewards, and environmental consequences.''

Use the title, message, and visual intent as semantic guidance, not as a fixed composition. Interpret them freely in a way that fits the topic and requested style. Choose any visual grammar, composition, and explanatory devices that communicate the idea clearly. No watermark.
\end{tcolorbox}

\section{Case Study and Further Discussion}
\label{sec:case-study-further-discussion}

This section presents five fixed-combination qualitative comparisons from the evaluation set.

The cases progress from a compact didactic brief to increasingly demanding narrative, causal, and dependency-rich briefs. In order, they cover spaced practice, collaborative animation, census-to-funding, multilingual classroom participation, and neighborhood readiness for extreme-weather outages.

Each figure compares the same five systems---\textsc{SlidesWeaver} (Topology First), \textsc{SlidesWeaver} (Text First), \textsc{AutoPresent}, \textsc{DeepSlides}, and \textsc{DeepPresenter}---on one case.

To ensure a fair comparison, we fix the language and image backbones across systems.
Here, \texttt{GPT+GPT} and \texttt{Qwen+GPT} pair GPT-5.4 and Qwen 3.8 Max, respectively, with GPT Image 2; the image-free baselines use the corresponding text backbone.

\subsection{Content Density Is Model-Dependent}

The comparisons expose a model-dependent content-density effect.
For some briefs, a language model expands explanations, adds qualifiers, or enumerates more steps than another model under the same slide budget.
The additional content can improve coverage, but it also increases the visual workload passed to planning and synthesis, producing denser text blocks and less negative space.
This effect is especially visible in cases whose success criterion depends on a multi-step explanation, such as the census-to-funding and emergency-readiness cases.
The observation motivates treating verbosity as a controllable generation variable.
Content planning should allocate detail jointly with the topology and the available visual area, instead of assuming that all language backbones produce comparable text volumes.

\subsection{Inpainting Quality Shapes Topology First Method}

The second pattern concerns spatial realization.
Topology First commits to relational structure before visual synthesis; its inpainting stage must honor registered regions and preserve ownership across neighboring elements.
When masks or boundaries are imprecise, foreground content can drift across regions and create overlap even when the topology is coherent.

This behavior is consistent with the synthesis ablation in Table~\ref{tab:ablation-synthesis}: layered generation raises Topology First Layout from $0.72$ to $0.87$ on average, whereas replacing layered generation with inpainting lowers Text First Layout from $0.95$ to $0.79$.
The case studies make the same dependency visible at the artifact level.
Topology planning supplies the structure, while image-model spatial controllability determines how faithfully that structure is rendered.

As suggested by the results in Table~\ref{tab:system-results}, more capable inpainting models provide a direct path to better Topology First decks.
Improved mask adherence and boundary control would preserve the route's freedom to explore expressive composition while reducing spill and overlap, yielding more elegant and polished editable presentations without changing the planning interface in the future.

\section{Evaluation Prompt and Scoring Procedure}
\label{app:evaluation-prompt}

We evaluate each generated deck with one blinded vision--language-model (VLM) request.
The request contains the frozen presentation brief (topic, audience, communication goal, success criterion, style brief, language, and requested slide count), deterministic generation evidence, the text extracted from the native PPTX, and an ordered montage of all requested slides.
The frozen judge is \texttt{gpt-5.6-terra}, queried at temperature~0 through the evaluation proxy.
The judge returns independent five-point scores for Visual Expressiveness (VE), Instruction Following (IF), and Content, together with a functional topology graph for eligible content slides and three serious-defect flags for every requested slide.
The judge never receives the system identity.

The remaining metrics are computed deterministically from the validated response and the generation manifest. 
First-Pass Success requires a successful and valid native artifact from outer attempt 1 with no failed slides. 
To provide the judge with a complete presentation whenever possible, a failed slide may be reconstructed for up to three additional attempts; such recovery does not alter its First-Pass Success status.

Layout is computed only over slides that are successfully generated and rendered in the evaluated artifact, consistent with Eq.~\ref{eq:layout}. For each such slide, Layout passes when the judge marks severe overlap, severe clipping, and severe out of bounds as false, and the deck-level score is the fraction of these successfully rendered slides that pass all three checks. Slides that remain failed after reconstruction are excluded from the Layout denominator and are instead reflected by the First-Pass Success metric.

Topology is emitted only for successfully rendered eligible content slides (2,\ldots,N-1), and its normalized two-iteration Weisfeiler--Lehman similarity is converted to Topo-Vendi. Generation cost and latency are summed from the event ledger and outer-attempt status files, while judge cost is logged separately.

The following is the original prompt used for the unified evaluation judge.
The prompt's output template is appended programmatically and validated before recording any metric.

\begin{tcolorbox}[
    breakable,
    colback=gray!5!white,
    colframe=black!75!black,
    title=Evaluation Prompts,
    fonttitle=\bfseries,
    boxrule=1pt,
    arc=3mm,
    left=6pt,right=6pt,top=6pt,bottom=6pt
]
\small
You are a strict blinded evaluator of one complete presentation. Evaluate only the presentation brief, deterministic evidence, extracted slide text, and ordered slide montage supplied in this request. Do not infer the generating system. Return five independent outputs: Visual Expressiveness (VE), Instruction Following (IF), Content, content-slide topology, and per-slide Layout flags.

\medskip
\textbf{Visual Expressiveness (VE).}
VE is an integer from 1 to 5. Judge purpose-built visual communication, information visualization, visual storytelling, visual richness, aesthetics, and requested-style match. Generic boxes, decorative color, repeated stock imagery, and text inside containers do not by themselves count as strong visual expression. Score 1 means visually broken or overwhelmingly text-only; 2 means sparse, generic, repetitive, or weak; 3 means adequate with noticeable limits; 4 means strong, varied, polished, and effective; 5 means exceptional purpose-built visual storytelling throughout. Do not let content correctness inflate VE.

\medskip
\textbf{Instruction Following (IF).}
IF is an integer from 1 to 5. Judge only compliance with explicit instructions: topic, target audience, communication goal, success criterion, style brief, language, requested slide count, and other explicit constraints. Use the deterministic generation evidence for slide count and failed pages. Score 1 means the instructions are mostly unmet; 3 means the main request is met with meaningful omissions; 5 means all explicit requirements are satisfied exceptionally well. Do not let visual polish excuse missed instructions.

\medskip
\textbf{Content.}
Content is an integer from 1 to 5 and combines content quality with narrative coherence. Judge relevance, specificity, informativeness, clarity, internal consistency, limited redundancy, logical slide-to-slide progression, appropriate emphasis, and whether the ending follows from the preceding material and serves the communication goal. Penalize generic filler, unsupported specificity, obvious factual errors, repetition, and disconnected slides. Score 1 means unusable or incoherent; 3 means adequate but generic, uneven, or weakly organized; 5 means exceptionally informative, clear, and coherent. Do not reward verbosity.

\medskip
\textbf{Content-slide topology.}
Analyze exactly the slide numbers listed under \texttt{ELIGIBLE CONTENT SLIDES}. Emit one topology object per eligible slide and no others. Regions may use only the types \texttt{title}, \texttt{text}, \texttt{visual}, \texttt{chart}, and \texttt{container}; size classes may be only \texttt{small}, \texttt{medium}, or \texttt{large}; groupings may be only \texttt{row}, \texttt{column}, \texttt{overlay}, \texttt{radial}, or \texttt{freeform}; and edges may use only \texttt{left\_of}, \texttt{above}, \texttt{adjacent}, \texttt{aligned}, \texttt{overlap}, or \texttt{contains}. Capture functional composition, not wording, color, image identity, or decorative detail. IDs must be unique within a slide, and every edge endpoint must reference a region ID from that slide.

\medskip
\textbf{Layout.}
Emit exactly one object for every requested slide. Mark only serious, visible defects: \texttt{severe\_overlap} is an unintended collision that materially obscures content; \texttt{severe\_clipping} is materially cut-off content; and \texttt{severe\_out\_of\_bounds} is content visibly extending beyond the canvas. Intentional overlays, text on shapes, and close spacing are not failures. A failed red-X placeholder is not itself an overlap, clipping, or out-of-bounds defect; deterministic code handles failed slides separately.

\medskip
Return exactly one JSON object matching \texttt{OUTPUT JSON TEMPLATE}. The topology and layout arrays in the template illustrate object shape only; populate the required slide numbers stated in the request. Do not return Markdown or additional keys.
\end{tcolorbox}

\section{Future Work}
Future work will explore two directions. First, the performance of Topology First is currently constrained by the spatial controllability and robustness of existing inpainting models, which may introduce visual drift or imperfect adherence to the grounded structure during synthesis. As inpainting models become more reliable in following masks, geometry, and regional semantics, we expect Topology First to better realize its potential for combining flexible composition with coherent visual integration. Second, given current resource constraints, we leave system-level post-training for future study. Incorporating post-training signals for presentation aesthetics and visual quality may further improve the polish, consistency, and overall visual appeal of generated slides~\cite{xie2026presentationintelligence}.

\begin{figure*}[t]
    \centering
    \includegraphics[width=0.85\textwidth]{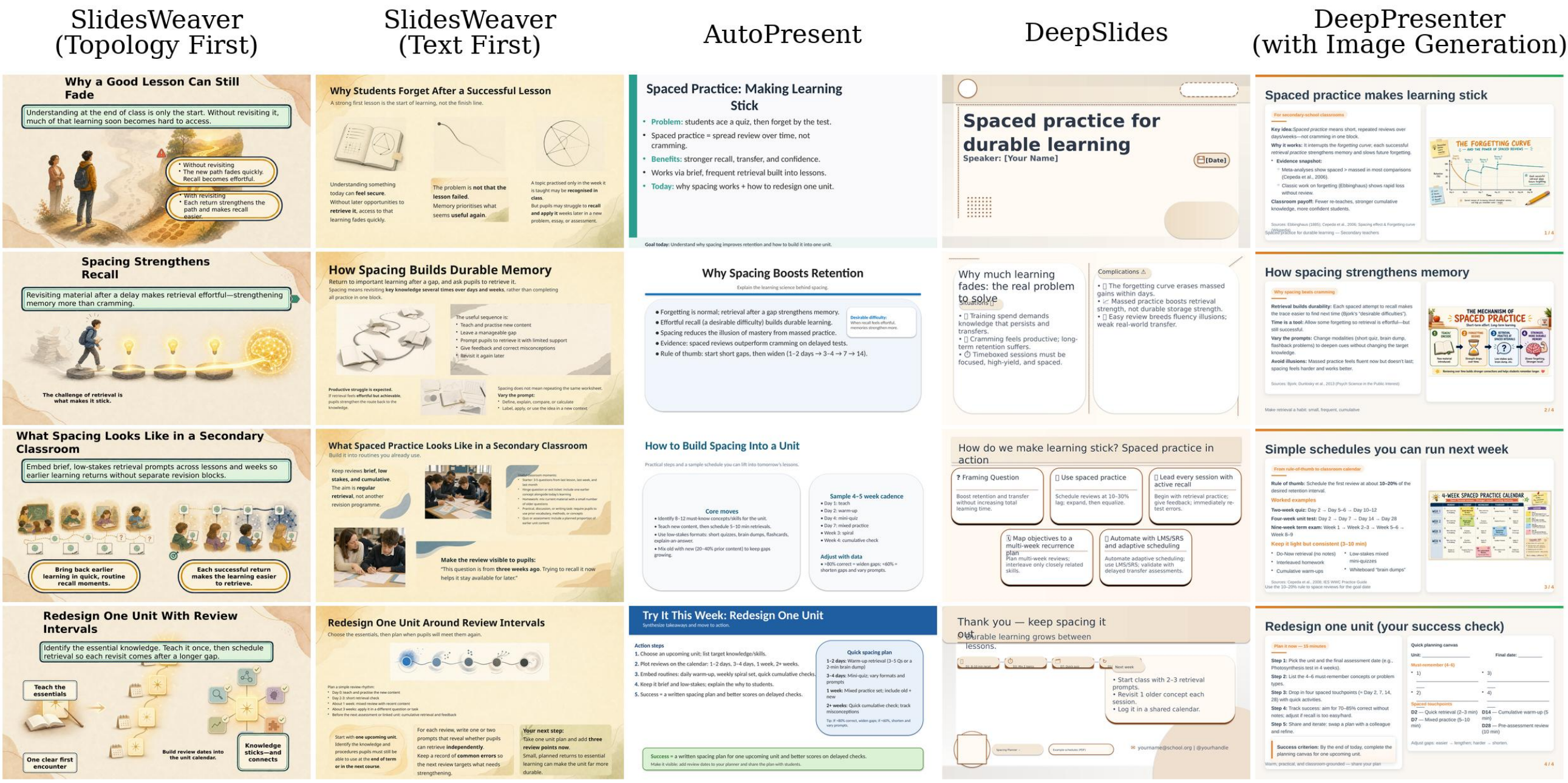}
    \caption{Fixed-combination comparison for ``Spaced practice for durable learning.'' This four-slide case uses \texttt{GPT+GPT} and provides the most compact brief in the set.}
    \label{fig:case-study-spaced-practice}
\end{figure*}

\begin{figure*}[t]
    \centering
    \includegraphics[width=0.85\textwidth]{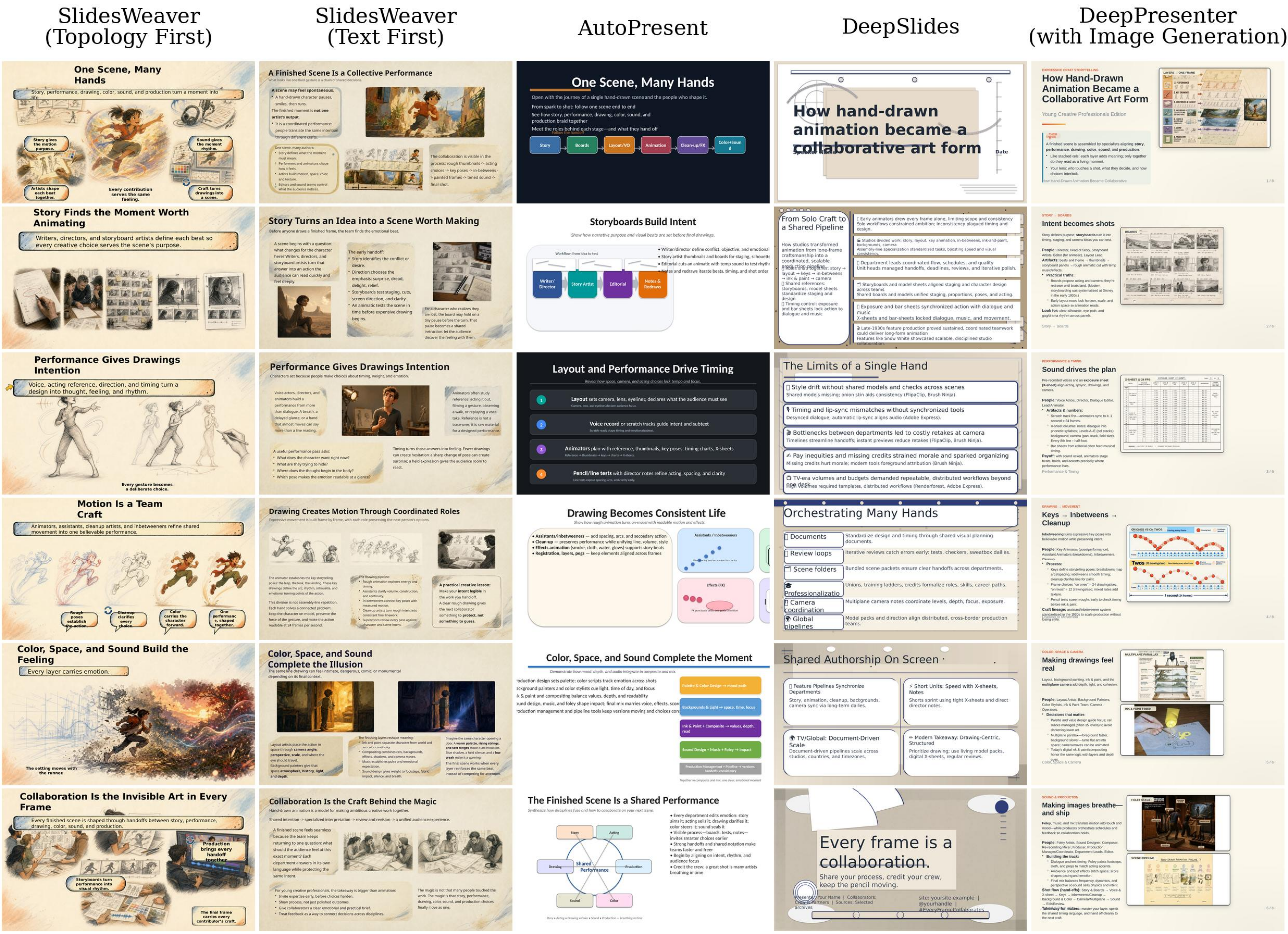}
    \caption{Fixed-combination comparison for ``How hand-drawn animation became a collaborative art form.'' This six-slide narrative case uses \texttt{GPT+GPT} and requires a coherent sequence of creative stages.}
    \label{fig:case-study-animation}
\end{figure*}

\begin{figure*}[t]
    \centering
    \includegraphics[width=0.85\textwidth]{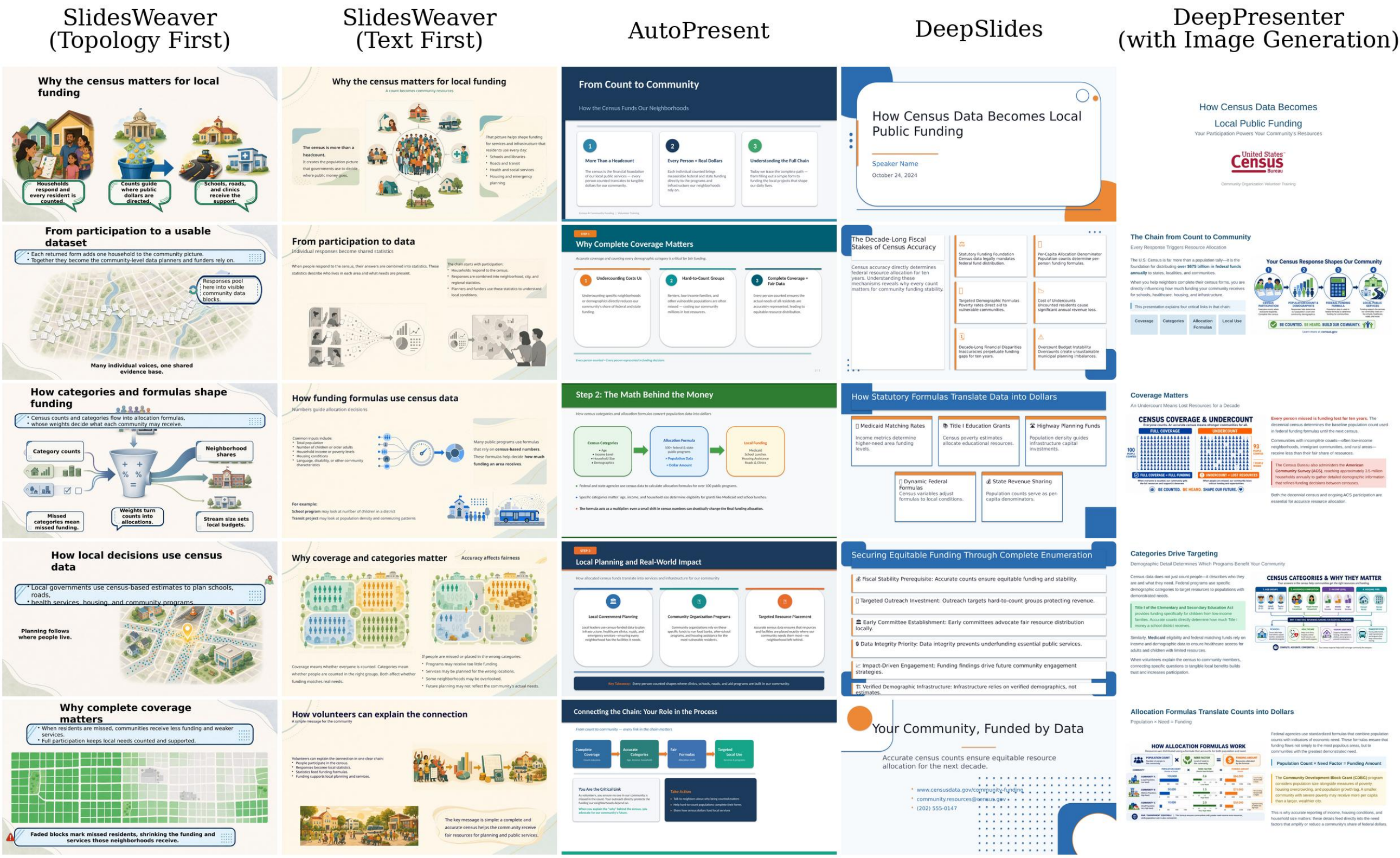}
    \caption{Fixed-combination comparison for ``How census data becomes local public funding.'' This goal-oriented case uses \texttt{Qwen+GPT} and combines a causal chain with an accessible public-information style.}
    \label{fig:case-study-census}
\end{figure*}

\begin{figure*}[t]
    \centering
    \includegraphics[width=0.85\textwidth]{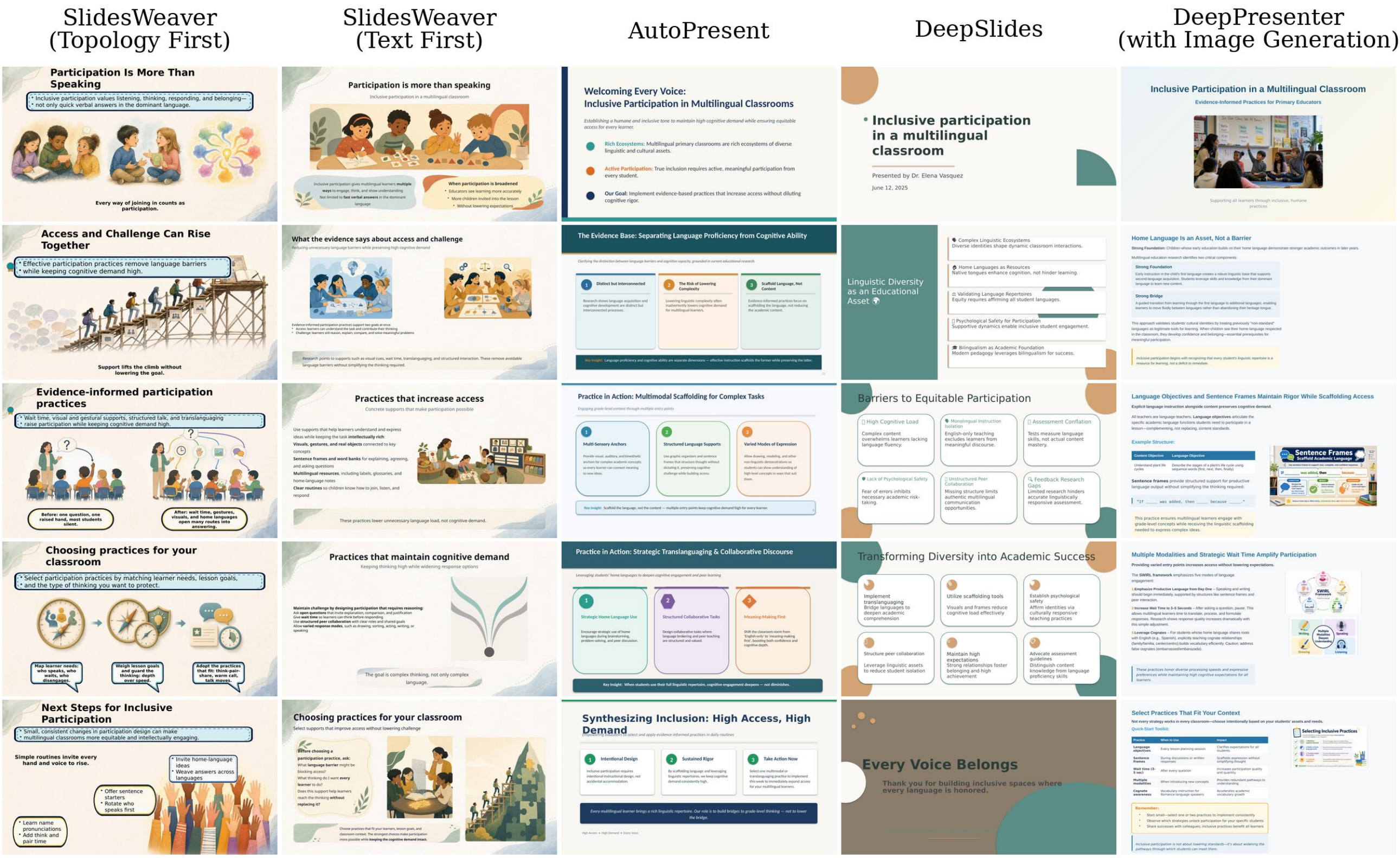}
    \caption{Fixed-combination comparison for ``Inclusive participation in a multilingual classroom.'' This high-demand case uses \texttt{Qwen+GPT} and must coordinate access, participation, and cognitive challenge.}
    \label{fig:case-study-multilingual}
\end{figure*}

\begin{figure*}[t]
    \centering
    \includegraphics[width=0.9\textwidth]{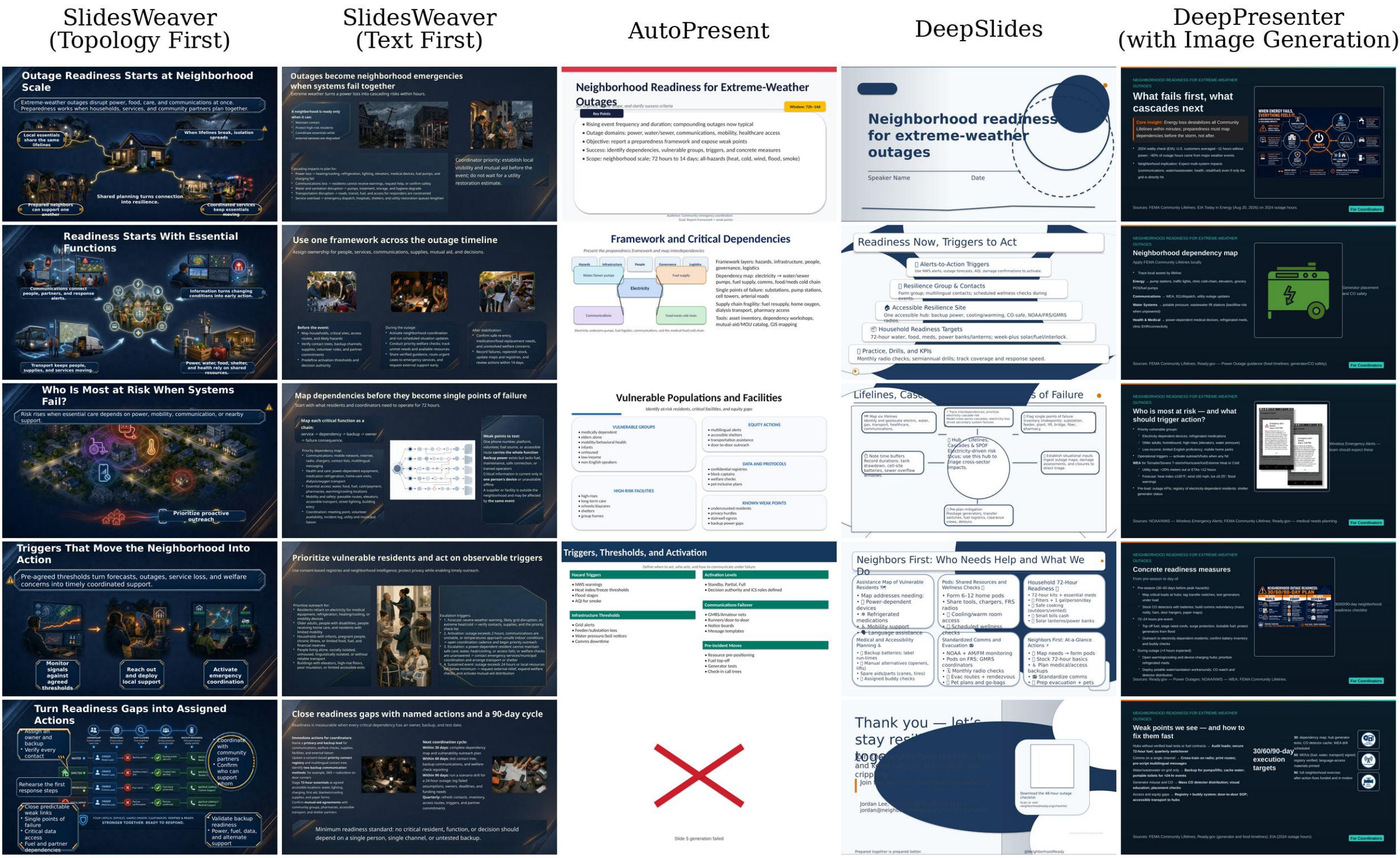}
    \caption{Fixed-combination comparison for ``Neighborhood readiness for extreme-weather outages.'' This high-demand case uses \texttt{GPT+GPT} and requires dependencies, vulnerable groups, triggers, and concrete readiness actions to remain visible together.}
    \label{fig:case-study-weather}
\end{figure*}
\end{document}